\documentclass[fleqn,usenatbib]{mnras}

\usepackage{newtxtext,newtxmath}

\usepackage[T1]{fontenc}
\usepackage{ae,aecompl}
\usepackage{hyperref}

\usepackage{graphicx}	
\usepackage{amsmath}	
\usepackage{amssymb}	
\usepackage{multirow}
\usepackage{float}
\usepackage{epsfig}
\hypersetup{draft}

\newcommand{\aips}{{$\cal AIPS\/$}}
\newcommand{\etal}{et~al.}
\def\emerlin{$e$-MERLIN}
\def\emerlinmerlin{($e$)MERLIN}
\def\emerge{$e$-MERGE}

\newcommand{\fsec}{\mbox{\ensuremath{.\!\!^{\rm s}}}}

\title[\emerge: Overview \& Survey Description]{The \emerlin\ Galaxy Evolution Survey (\emerge): Overview and Survey Description}

\author[T.\,W.\,B.\ Muxlow et al.]{T.\,W.\,B. Muxlow$^{1}$,
A.\,P. Thomson$^{1,2}$\thanks{E-mail: alasdair.thomson@manchester.ac.uk},
J.\,F. Radcliffe$^{1,3}$,
N.\,H. Wrigley$^{1}$,
R.\,J. Beswick$^{1}$,\newauthor
Ian Smail$^{4}$,
I.\,M.\ M$\rm^{c}$Hardy$^{2}$,
S.\,T.\ Garrington$^{1}$,
R.\,J.\ Ivison$^{5,6}$,
M.\,J.\ Jarvis$^{7,8}$,\newauthor
I. Prandoni$^{9}$,
M.\ Bondi$^{9}$,
D.\ Guidetti$^{9}$,
M.\,K.\ Argo$^{10}$,
David Bacon$^{11}$,
P.\,N.\ Best$^{6}$,\newauthor
A.\,D.\ Biggs$^{5}$,
S.\,C.\ Chapman$^{12}$,
K.\ Coppin$^{13}$,
H.\ Chen$^{1,14,15}$,
T.\,K.\ Garratt$^{13}$,\newauthor
M.\,A.\ Garrett$^{1,16}$,
E.\ Ibar$^{17}$,
Jean-Paul\ Kneib$^{18,19}$,
Kirsten K.\ Knudsen$^{20}$,\newauthor
L.\,V.\,E.\ Koopmans$^{21}$,
L.\,K.\ Morabito$^{4}$,
E.\,J.\ Murphy$^{22}$,
A.\ Njeri$^{1}$,
Chris Pearson$^{23}$,\newauthor
M.\,A.\ P\'{e}rez-Torres$^{24}$,
A.\,M.\,S.\ Richards$^{1}$,
H.\,J.\,A.\ R\"{o}ttgering$^{16}$,
M.\,T.\ Sargent$^{25}$,\newauthor
Stephen Serjeant$^{26}$,
C.\ Simpson$^{27}$,
J.\,M.\ Simpson$^{28}$,
A.\,M.\ Swinbank$^{4}$,
E.\ Varenius$^{20,1}$,\newauthor
T.\ Venturi$^{9}$
\\
\parbox{\textwidth}{
\vspace*{4mm}
Author affiliations shown in Appendix\,\ref{appendix:affiliations}
 }
}

\date{Accepted 2020 May 5; Received 2020 May 5; in original form 2020 February 29}

\pubyear{2020}

\begin{document}
\label{firstpage}
\pagerange{\pageref{firstpage}--\pageref{lastpage}}
\maketitle

\begin{abstract}
We present an overview and description of the \emerlin\ Galaxy Evolution survey (\emerge) Data Release 1 (DR1), a large program of high-resolution 1.5\,GHz radio observations of the GOODS-N field comprising $\sim140$\,hours of observations with \emerlin\ and $\sim40$\,hours with the Very Large Array (VLA). We combine the long baselines of \emerlin\ (providing high angular resolution) with the relatively closely-packed antennas of the VLA (providing excellent surface brightness sensitivity) to produce a deep 1.5\,GHz radio survey with the sensitivity ($\sim 1.5\mu$Jy\,beam$^{-1}$), angular resolution ($0\farcs2$--$0\farcs7$) and field-of-view ($\sim15\arcmin\times 15\arcmin$) to detect and spatially resolve star-forming galaxies and AGN at $z\gtrsim 1$. The goal of \emerge\ is to provide new constraints on the deep, sub-arcsecond radio sky which will be surveyed by SKA1-mid. In this initial publication, we discuss our data analysis techniques, including steps taken to model in-beam source variability over a $\sim20$\,year baseline and the development of new point spread function/primary beam models to seamlessly merge \emerlin\ and VLA data in the $uv$ plane. We present early science results, including measurements of the luminosities and/or linear sizes of $\sim500$\ galaxes selected at 1.5\,GHz. In combination with deep Hubble Space Telescope observations, we measure a mean radio-to-optical size ratio of $r_{\rm eMERGE}/r_{\rm HST}\sim1.02\pm0.03$, suggesting that in most high-redshift galaxies, the $\sim$GHz continuum emission traces the stellar light seen in optical imaging. This is the first in a series of papers which will explore the $\sim$kpc-scale radio properties of star-forming galaxies and AGN in the GOODS-N field observed by \emerge\ DR1.
\end{abstract}




\section{Introduction}

Historically, optical and near-infrared surveys have played a leading role in measuring the integrated star formation history of the Universe \citep[e.g.][]{lilly96,madau96}, however in recent years a pan-chromatic (i.e. X-ray -- radio) approach has become key to achieving a consensus view on galaxy evolution \citep[e.g.][]{scoville07,driver09}. Since the pioneering work in the far-infrared (FIR) and sub-millimetre wavebands undertaken with the Submillimeter Common-User Bolometer Array (SCUBA) on the James Clerk Maxwell Telescope (JCMT), it has been established that a significant fraction of the integrated cosmic star formation \citep[up to $\sim 50\%$ at $z\sim1$--$3$;][]{swinbank14, barger17} has taken place in heavily dust-obscured environments, which can be difficult (or impossible) to measure fully with even the deepest optical/near-infrared data \citep[e.g.][]{barger98,seymour08,hodge13,casey14}. Within this context, deep interferometric radio continuum observations are an invaluable complement to studies in other wavebands, providing a dust-unbiased tracer of star formation \citep[e.g.][]{condon92,smolcic09}, allowing us to track the build-up of stellar populations through cosmic time without the need to rely on uncertain extinction corrections. Moreover, radio continuum observations also provide a direct probe of the synchrotron emission produced by active galactic nuclei (AGN), which are believed to play a crucial role in the evolution of their host galaxies via feedback effects \citep{best06, schaye15, harrison18}.

The radio spectra of galaxies at $\gtrsim 1$\,GHz frequencies are typically thought to result from the sum of two power-law components \citep[e.g.][]{condon92, murphy11}.  At frequencies between $\nu_{\rm rest}\sim 1$--$10$\,GHz, radio observations trace steep-spectrum ($\alpha\sim-0.8$, where $S_{\nu}\propto\nu^\alpha$) synchrotron emission, which can be produced either by supernova explosions \citep[in which case it serves as a dust-unbiased indicator of the star-formation rate, SFR, over the past $\sim 10$--$100$\,Myr: ][]{bressan02} or from accretion processes associated with the supermassive black holes (SMBHs) at the centres of AGN hosts. At higher frequencies ($\nu_{\rm rest}\gtrsim 10$\,GHz), radio observations trace flatter-spectrum ($\alpha\sim-0.1$) thermal free-free emission, which signposts the scattering of free-electrons in ionised H{\sc ii} regions around young, massive stars, and thus is considered to be an excellent tracer of the \textit{instantaneous} SFR. 

This dual origin for the radio emission in galaxies (i.e. star-formation and AGN activity) makes the interpretation of monochromatic radio observations of unresolved, distant galaxies non-trivial. To determine the origin of radio emission in distant galaxies requires (a) the angular resolution and surface brightness sensitivity to morphologically decompose (extended) star-formation and radio jets from (point-like) nuclear activity \citep[e.g.][]{baldi18, jarvis19}, and/or (b) multi-frequency observations which provide the spectral index information necessary to measure reliable rest-frame radio luminosities. These allow galaxies which deviate from the FIR/radio correlation (FIRRC) to be identified, a correlation on which star-forming galaxies at low and high-redshift are found to lie \citep[e.g.][]{helou85, bell03, ivison10, thomson14, magnelli15}. 

\begin{figure*}
\centering
\includegraphics[width=\linewidth]{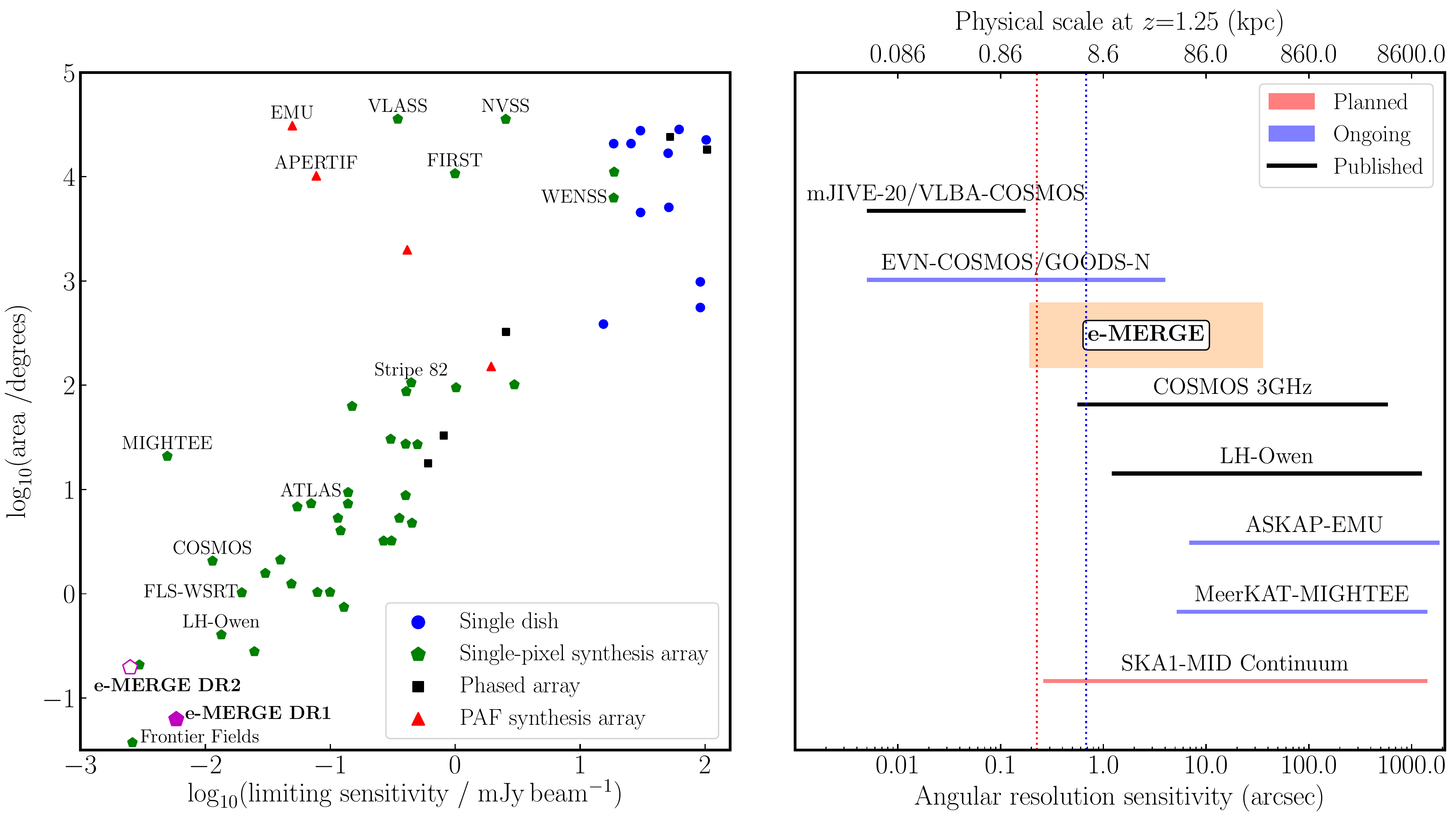}
\caption{\textit{Left:} Sky area versus sensitivity (detection limit or $5\sigma_{\rm rms}$) for selected radio surveys, highlighting the sensitivity of \emerge\ Data Release 1 with respect to existing studies in the $\sim$GHz window. In a forthcoming Data Release 2, including $\sim 4\times$ more \emerlin\ $uv$ data, we will quadruple the area and double the sensitivity of \emerge\, offering the first sub-$\mu$Jy\,beam$^{-1}$ view of the deep 1.5\,GHz radio sky. \textit{Right:} A comparison of the angular scales probed by selected $\sim$GHz-frequency radio continuum surveys; the right-most edge of each line represents the Largest Angular Scale ($\theta_{\rm LAS}$) probed by the corresponding survey, and is defined by the shortest antenna spacing in the relevant telescope array. The left-most edge is the angular resolution ($\theta_{\rm res}$) defined by the naturally-weighted PSF of each survey. Vertical lines at $0\farcs25$ and $0\farcs70$ (corresponding to $\sim2$\,kpc and $\sim7$\,kpc at $z=1.25$, respectively) represent the typical effective radii of massive ($M_\star \sim 10^{11}$\,M$_\odot$) early- and late-type galaxies seen in optical studies \citep{vanderwel14}. While the area coverage of \emerge\ DR1 is modest compared with other surveys, its combination of high sensitivity and sub-arcsecond angular resolution offers a unique view of the population of radio-selected SFGs and AGN at high redshift. The long baselines of \emerlin\ bridge the gap between VLA and Very Long Baseline Interferometry (VLBI) surveys, offering sensitive imaging at $\sim$kpc scale resolution in the high-redshift Universe. \emerge\ thus provides a crucial benchmark for the sizes and morphologies of the high redshift radio source population, and delivers a glimpse of the radio sky that will be studied by SKA1-mid in the next decade.}\label{fig:e-MERGE_ang_res}
\end{figure*}

The magnification afforded by gravitational lensing provides one route towards probing the obscured star-formation and AGN activity via radio emission in individual galaxies at high-redshift \citep[e.g.][]{hodge15, thomson15}, however in order to produce a statistically-robust picture of the interplay between these processes for the high-redshift galaxy population \textit{in general}, and to obtain unequivocal radio counterparts for close merging systems requires sensitive ($\sigma_{\rm rms}\sim1\,\mu$Jy\,beam$^{-1}$) radio imaging over representative areas ($\gtrsim10\arcmin\times 10\arcmin$) with $\sim$kpc (i.e. sub-arcsecond) resolution. The Karl G.\,Jansky Very Large Array (VLA) is currently capable of delivering this combination of observing goals in $S$-band (3\,GHz), $X$-band (10\,GHz), and at higher frequencies. However by $z\sim2$ these observations probe rest-frame frequencies $\nu_{\rm rest}\gtrsim 10$--$30$\,GHz, a region of the radio spectrum in which the effects of spectral curvature may become important due to the increasing thermal free-free component at high-frequencies \citep[e.g.][]{murphy11}, and/or spectral steepening due to cosmic ray effects \citep{galvin18, thomson19b} and free-free absorption \citep{tisanic19}. This potential for spectral curvature complicates efforts to measure the rest-frame radio luminosities (conventionally, $L_{\rm 1.4\,GHz}$) of high redshift galaxies from these higher-frequency observations.

\begin{figure*}
\centering
\includegraphics[width=0.9\linewidth]{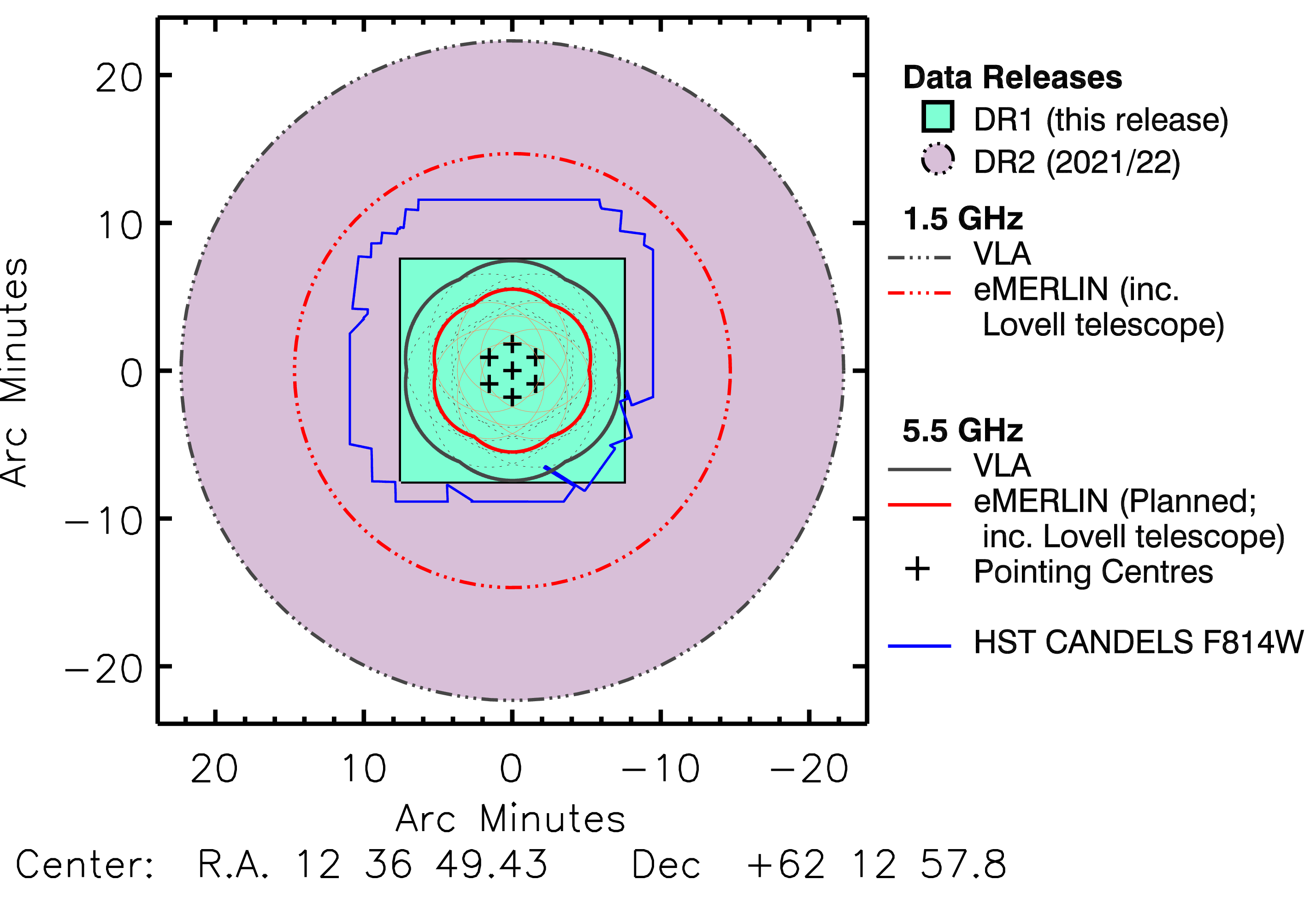}
\caption{The eMERGE survey layout, showing the current (DR1; black box) and planned future (DR2; lilac circle) survey areas. \emerge\ 1.5\,GHz observations comprise a single deep pointing which includes 40\,hours of VLA and 140\,hours of \emerlin\ observations, encompassing the HST CANDELS field (shown in blue). Our DR1 area is limited by time and bandwidth smearing effects (both of which increase as a function of radial distance from the phase centre: see \S\,\ref{sect:smearing} for details). In a forthcoming DR2, we will include an additional $\sim 400$\,hours of observed \emerlin\ 1.5\,GHz data, which will be processed without averaging in order to allow the full primary beam of the 25\,m \emerlin\ and VLA antennas to be mapped. \emerge\ DR1 includes the 14\,h seven-pointing 5.5\,GHz VLA mosaic image published by \citet{Guidetti2017:em}, which will be supplemented in our forthcoming DR2 with an additional 42\,hours of VLA and $\sim 380$\,hours of \emerlin\ 5.5\,GHz observations which share the same pointing centres. Our planned 5.5\,GHz mosaic will eventually reach an angular resolution of $\sim 50$\,mas at $\sigma_{\rm 5.5\,GHz}\sim0.5\,\mu$Jy\,beam$^{-1}$. Note that the VLA 5.5\,GHz pointings are significantly over-sampled with respect to the VLA primary beam in order to facilitate $uv$ plane combination with data from \emerlin, whose primary beam is significantly smaller than the VLA's when the 76\,m Lovell telescope is included in the array.}\label{fig:eMERGE FoV}
\end{figure*}

Furthermore, the instantaneous field of view (FoV) of an interferometer is limited by the primary beam, $\theta_{\rm PB}$, which scales as $\lambda/D$, with $D$ being the representative antenna diameter. At 1.4\,GHz the FoV of the VLA's 25\,m antennas is $\theta_{\rm PB}\sim32'$, while the angular resolution offered by its relatively compact baselines ($B_{\rm max}=36.4$\,km) is $\theta_{\rm res}\sim 1\farcs5$. This corresponds to $\sim 12$\,kpc at $z\sim 2$, and is therefore insufficient to morphologically study the bulk of the high-redshift galaxy population, which have optical sizes of only a few kpc \citep{vanderwel14}. At 10\,GHz, in contrast, the angular resolution of the VLA is $\theta_{\rm res}\sim 0\farcs2$ ($\sim1.5$\,kpc at $z=2$), but the FoV shrinks to $\theta_{\rm PB}\sim 4\farcm5$. This large (a factor $\sim50\times$) reduction in the primary beam area greatly increases the cost of surveying deep fields over enough area to overcome cosmic variance \citep[e.g][]{murphy17}, particularly given that the positive \textit{k}-correction in the radio bands means that these observations probe an intrinsically fainter region of the rest-frame radio SEDs of high-redshift galaxies to begin with. 

Over the coming decade the SKA1-mid and its precursor instruments (including MeerKAT and ASKAP) will add new capabilities to allow the investigation of the faint extragalactic radio sky \citep{prandoni15,jarvis2016,taylor2017}. At $\sim$1\,GHz observing frequencies these extremely sensitive instruments will reach (confusion-limited) $\sim\mu$Jy\,beam$^{-1}$ sensitivities over tens of square degrees in area, but with an angular resolution of $\gtrsim10$\,arcseconds, corresponding to a linear resolution of $\gtrsim 80$\,kpc at $z=1$. Crucially, this means that a significant fraction of the high-redshift star-forming galaxies and AGN detected in these surveys will remain unresolved (see Fig.\,\ref{fig:e-MERGE_ang_res}).

There is thus a need for high angular resolution and high sensitivity, wide-field radio observations in the $\sim$GHz radio window to complement surveys which are underway in different frequency bands, and with different facilities. To address this, we have been conducting a multi-tiered survey of the extragalactic sky using the \textit{enhanced} Multi-Element Remotely Linked Interferometer Network (\emerlin), the UK's national facility for high angular resolution radio astronomy (Garrington \etal, in prep), along with observations taken with the VLA. This ongoing project -- the \emerlin\ Galaxy Evolution Survey (eMERGE) -- exploits the unique combination of the high angular resolution and large collecting area of \emerlin, and the excellent surface brightness sensitivity of the VLA. The combination of these two radio telescopes allows the production of radio maps which exceed the specifications of either instrument individually, and thus allows synchrotron emission due to both star-formation activity and AGN to be mapped in the high-redshift Universe.

\subsection{\emerge: an \emerlin\ legacy project}

\emerlin\ is an array of seven radio telescopes spread across the UK (having a maximum baseline length $B_{\rm max}=217$\,km), with antenna stations connected via optical fibre links to the correlator at Jodrell Bank Observatory. \emerlin\ is an inhomogeneous array comprised of the 76\,m Lovell Telescope at Jodrell Bank (which provides $\sim 58\%$ of the total \emerlin\ collecting area), one 32\,m antenna near Cambridge (which provides the longest baselines) and five 25\,m antennas, three of which are identical in design to those used by the VLA. 

Due to the inhomogeneity of the \emerlin\ telescopes, the primary beam response (which defines the sensitivity of the array to emission as a function of radial distance from the pointing centre) is complicated (see \S\,\ref{sect:primarybeam}), however to first order it can be parameterised at 1.5\,GHz as a sensitive central region $\sim 15\arcmin$ in diameter (arising from baselines which include the Lovell Telescope) surrounded by a $\sim45\arcmin$ annulus, which is a factor $\sim 2\times$ less sensitive, and arises from baselines between pairs of smaller telescopes. 

Our target field for \emerge\ is the Great Observatories Origins Deep Survey North field \citep[GOODS-N, $\alpha=12^{\rm h}36^{\rm m}49\fsec40$, $\delta=+62^{\circ}12^{\prime}58\farcs0$;\ ][]{dickinson03}, which contains the original Hubble Deep Field \citep{williams96}. Due to the extent of the deep multi-wavelength coverage, GOODS-N remains one of the premier deep extra-galactic survey fields. The field was first observed at $\sim1.4$\,GHz (L-band) radio frequencies by the VLA by \citet{richards00}, yielding constraints on the $\sim 10$--$100\,\mu$Jy radio source counts. Using a sample of 371 sources, \citet{richards00} found flattening of the source counts (normalised to $N(S)\propto S^{3/2}$) below $S_{\rm 1.4\,GHz}=100\,\mu$Jy. Later, \citet{morrison10}, using the original \citet{richards00} observations plus a further 121 hours of (pre-upgrade) VLA observations achieved improved constraints on the radio source counts, finding them to be nearly Euclidian at flux densities $\lesssim 100\,\mu$Jy and with a median source diameter of $\sim 1\farcs2$, i.e.\ close to the angular resolution limit of the VLA. \citet{Muxlow2005:hdf} subsequently published 140\,hours of 1.4\,GHz observations of GOODS-N with MERLIN, obtaining high angular resolution postage stamp images of 92 of the \citet{richards00} VLA sources, a slight majority of which (55/92) were found to be associated with \textit{Chandra} X-ray sources \citep{brandt2001,richards2007}, and hence were classified as possible AGN. The angular size distribution of these bright radio sources peaks around a largest angular scale of $\theta_{\rm LAS}\sim1\farcs0$, but with tail of more extended sources out to $\theta_{\rm LAS}\sim4\farcs0$.

More recently, the field has been re-observed with the upgraded VLA by \citet{owen18}, who extracted a catalogue of 795 radio sources over the inner $\sim9\arcmin$ of the field. \citet{owen18} measured a linear size distribution in the radio which peaks at $\sim 10$\,kpc, finding the radio emission in most galaxies to be larger than the galaxy nucleus but smaller than the galaxy optical isophotal size ($\sim 15$--$20$\,kpc).

In this paper, we present a description of our updated \emerlin\ observations of the field, which along with an independent reduction of the \citet{owen18} VLA observations and older VLA/MERLIN observations, constitute \emerge\ Data Release 1 (DR1). This data release will include $\sim 1/4$ of the total \emerlin\ L-Band (1--2\,GHz) observations granted to the project (i.e.\ 140 of 560\,hours), which use the same pointing centre as all the previous deep studies of the field discussed in the preceding paragraphs. We use VLA observations to fill the inner portion of the $uv$ plane, which is not well-sampled by \emerlin, in order to enhance our sensitivity to emission on $\gtrsim 1\arcsec$ scales. We compare the survey area, sensitivity and angular resolution of \emerge\ with those of other state-of-the-art deep, extragalactic radio surveys in Fig.\,\ref{fig:e-MERGE_ang_res}. In addition to our L-Band observations, \emerge\ DR1 includes the 7-pointing VLA C-Band (5.5\,GHz) mosaic image previously published by \citet{Guidetti2017:em}. We summarise our \emerge\ DR1 observations in Table\,\ref{tab:e-MERGE-Data-products}, list the central coordinates of each \emerge\ pointing (1.5\,GHz and 5.5\,GHz using both telescopes) in Table\,\ref{table:eMERGE pointings}, and show the \emerge\ survey footprint (including both existing and planned future observations) in Fig.\,\ref{fig:eMERGE FoV}.

We describe the design, execution and data reduction strategies of \emerge\ DR1 in detail in \S\,\ref{sect:data-all}, including a discussion of the wide-field imaging techniques which we have developed to combine and image our \emerlin\ and VLA observations in \S\,\ref{sect:imaging}. We present early science results from \emerge\ DR1 in \S\,\ref{sect:results}, including the luminosity-redshift plane and angular size distribution of $\sim500$ high-redshift SFGs/AGN ($\sim250$ of which benefit from high-quality photometric redshift information from the literature), and demonstrate the image quality via a brief study of a representative $z=1.2$ submillimetre-selected galaxy (SMG) selected from our wide-field ($\theta_{\rm PB}=15\arcmin$), sensitive ($\sim 2\,\mu$Jy\,beam$^{-1}$), high-resolution ($\theta_{\rm res}\sim 0\farcs5$) 1.5\,GHz imaging of the GOODS-N field\footnote{\emerge\ is an \emerlin\ legacy survey, and therefore exists to produce lasting legacy data and images for the whole astronomical community. An \emerge\ DR1 source catalogue will be released in a forthcoming publication. After a short proprietary period, the full suite of \emerge\ DR1 wide-field images will be made available to the community. We encourage potential external collaborators and other interested parties to visit the \emerge\ website for the latest information: \url{http://www.e-merlin.ac.uk/legacy-emerge.html}}. Finally, we summarise our progress so far and outline our plans for future science delivery from \emerge\ (including the delivery of the full DR1 source catalogue) in \S\,\ref{sect:conclusions}. Throughout this paper we use a Planck\,2018 Cosmology with $H_0=67.4$\,km\,s$^{-1}$\,Mpc$^{-1}$ and $\Omega_m = 0.315$ \citep{planck18}.

\section{Observations \& Data Reduction}\label{sect:data-all}

\subsection{\textit{e}-MERLIN 1.5\,GHz}\label{sect:emerlin-dr}

The cornerstone of \emerge\ DR1 is our high-sensitivity, high-resolution L-band (1.25-1.75\,GHz; central frequency of 1.5\,GHz) imaging of the GOODS-N field, which we observed with \emerlin\ in five epochs between 2013\,Mar -- 2015\,Jul (a total on-source time of 140\,hours). In the standard observing mode, these \emerlin\ observations yielded time resolution of 1\,s/integration and frequency resolution of 0.125\,MHz/channel. The \emerlin\ frequency coverage is comprised of eight spectral windows (spws) with 512\,channels per spw per polarisation. We calibrated the flux density scale using $\sim$30\,minute scans of 3C\,286 at the beginning of each run, and tracked the complex antenna gains using regular $\sim 5$\,min scans of the bright phase reference source J1241+6020, which we interleaved between 10\,min scans on the target field. We solved for the bandpass response of each observation using a $\sim$30\,minute scan of the standard \emerlin\ L-band bandpass calibration source, OQ\,208 (1407+284). After importing the raw telescope data in to the NRAO Astronomical Image Processing System \citep[\aips:][]{Griesen2003:aips}, we performed initial \textit{a priori} flagging of known bad data -- including scans affected by hardware issues and channel ranges known to suffer from persistent severe radio frequency interference (RFI) -- using the automated {\sc serpent} tool \citep{peck13}, before averaging the data by a factor $4\times$ in frequency (to 0.5\,MHz resolution) in order to reduce the data volume, using the \aips\ task {\sc splat}. The discretisation of interferometer $uv$ data in time and frequency results in imprecisions in the ($u$,$v$) coordinates assigned to visibilities, which inevitably induces ``smearing'' effects in the image plane: the effect of this frequency averaging on the image fidelity will be discussed in \S\,\ref{sect:smearing}.

Next, we performed a further round of automated flagging to excise bad data, before further extensive manual flagging of residual time-variable and low-level RFI was carried out.

\begin{table*}
\begin{centering}
\caption[Summary of observations included within \emerge\ Data Release-1 (DR1).]{Summary of observations included within \emerge\ Data Release-1 (DR1).}
\label{tab:e-MERGE-Data-products}
\begin{tabular}{lccccccccccc}
\hline
\multicolumn{1}{l}{Telescope} &
\multicolumn{1}{c}{Reference} & 
\multicolumn{1}{c}{Array} &
\multicolumn{1}{c}{Project} &
\multicolumn{1}{c}{Total time} &
\multicolumn{1}{c}{Epoch(s)} & 
\multicolumn{1}{c}{Typical sensitivity} \\
\multicolumn{1}{l}{} &
\multicolumn{1}{c}{Frequency} &
\multicolumn{1}{c}{Config.} &
\multicolumn{1}{c}{Code} &
\multicolumn{1}{c}{(hours)} &
\multicolumn{1}{c}{} &
\multicolumn{1}{c}{($\mu$Jy\,beam$^{-1}$)} &\\
\hline
\emerlin\ $^1$ & 1.5\,GHz & -- & LE1015 & 140 & 2013 Mar \& Apr, 2013 Dec, 2015 Jul & 2.81\\
VLA$^2$ & 1.5\,GHz & A & TLOW0001 & 38 & 2011 Aug \& Sep & 2.01\\
MERLIN$^3$ & 1.4\,GHz & -- & -- & 140 & 1996 Feb -- 1997 Sep & 5.70\\
VLA$^{3,4}$ & 1.4\,GHz & A & -- & 42 & 1997 Sep -- 2000 May & 7.31\\
\hline
VLA$^{5,*}$ & 5.5\,GHz & B & 13B-152 &  2.5 & 2013 Sep & 7.90 \\
VLA$^{5,*}$ & 5.5\,GHz & A & 12B-181 &  14 & 2012 Oct & 3.22 \\

\hline
\end{tabular}
\end{centering}
\\ {\small References: 
$^1$ this paper;
$^2$Data originally presented by \citet{owen18}, but re-reduced in this paper;
$^3$\citet{Muxlow2005:hdf};
$^4$ \citet{Richards1998};
$^5$ \citet{Guidetti2017:em}.
$^*$ Observations comprise a seven-pointing mosaic.}
\end{table*}

\begin{table}
\caption{Pointing centres for the eMERGE observations. The same positions are (or will be) used for both VLA and \emerlin\ observations at a given frequency.}
\begin{centering}
\begin{tabular}{ccc}
\hline 
Band & R.A. & Dec. \\
& [hms (J2000)] & [dms (J2000)]\\
\hline 
\multirow{1}{*}{L (1.5\,GHz)} & $12^{\rm h}36^{\rm m}49\fsec40$ & $+62^{\circ}12^{\prime}58\farcs0$\\
\hline
\multirow{6}{*}{C (5.5\,GHz)} & $12^{\rm h}36^{\rm m}49\fsec40$ & $+62^{\circ}12^{\prime}58\farcs0$ \\
&$12^{\rm h}36^{\rm m}49\fs40$ & $+62^{\circ}14^{\prime}46\farcs0$\\
&$12^{\rm h}36^{\rm m}36\fsec00$ & $+62^{\circ}13^{\prime}52\farcs0$\\
&$12^{\rm h}36^{\rm m}36\fsec00$ & $+62^{\circ}12^{\prime}02\farcs0$\\
&$12^{\rm h}36^{\rm m}49\fsec40$ & $+62^{\circ}11^{\prime}10\farcs0$\\
&$12^{\rm h}37^{\rm m}02\fsec78$ & $+62^{\circ}12^{\prime}02\farcs0$\\
&$12^{\rm h}37^{\rm m}02\fsec78$ & $+62^{\circ}13^{\prime}52\farcs0$\\
\hline
\end{tabular}
\par\end{centering}
\label{table:eMERGE pointings}
\end{table}

\subsubsection{Amplitude calibration \& phase referencing}

We set the flux density scale for our observations using a model of 3C\,286 along with the flux density measured by \citet{PerleyButler2013:fl}.

The delays and phase corrections were determined using a solution interval matching the calibrator scan lengths. Any significant outliers were identified and removed. Initial phase calibration was performed for the flux calibrator using a model of the source, and for the phase and bandpass calibrators assuming point source models. These solutions were applied to all sources and initial bandpass corrections (not including the intrinsic spectral index of OQ\,208) were derived. The complex gains (phase and amplitude) were iteratively refined, with solutions inspected for significant outliers after each iteration to identify and exclude residual low level RFI before the complex gain calibration was repeated.

The solution table containing the complex gains was used to perform an initial bootstrapping of the flux density from 3C\,286 to the phase and bandpass calibrator sources. Exploiting the large fractional bandwidth of \emerlin\ ($\Delta\nu/\nu\sim 0.33$), these bootstrapped flux density estimates were subsequently improved by fitting the observed flux densities for J1241+6020 and OQ\,208 linearly across all eight spws. 

With the flux density scale and the spectral indices of the phase and bandpass calibrators thus derived, the bandpass calibration was improved, incorporating the intrinsic source spectral index. The complex gains were improved and then applied to all sources, including the target field. Finally, the target field was split from the multi-source dataset and the data weights were optimised based on the post-calibration baseline rms noise.

\subsubsection{Self-calibration}

We identified the brightest 26 sources ($S_{\rm 1.5\,GHz}\geq 120\,\mu$Jy) in the GOODS-N field at 1.5\,GHz \citep[guided by the catalogue of][]{Muxlow2005:hdf} and produced \emerlin\ thumbnail images over a $5\arcsec\times 5\arcsec$ region centred on each source. The sky model generated from these thumbnail images was used to produce a multi-source model for phase-only self-calibration. This used a solution interval equal to the scan duration and was repeated until the phase solution converged to zero (typically within $\sim 3$ iterations per epoch of data).

\subsubsection{Variability, flux density and astrometric cross-checks}\label{sect:variability}

Previous studies have shown that the fraction of sub-$100\mu$Jy variable radio sources is low \citep[a few percent, e.g.][]{mooley2016,radcliffe19}. However, relatively small levels of intrinsic flux density variability of sources in the field, along with any small discrepancies in the relative flux density scale assigned to each epoch, will result in errors in the final combined image if not properly accounted for.   

In order to assess and mitigate the effect of intrinsic source variability in our final, multi-epoch dataset, each epoch of \emerlin\ and VLA data was imaged and catalogued separately using the flood-filling algorithm \texttt{BLOBCAT} \citep{Hales2012:blo}, using rms maps generated by the accompanying \texttt{BANE} software \citep{Hancock2018:ae}. We cross-checked the catalogues from each epoch to identify sources with significant intrinsic variability ($\gtrsim 15\%$; greater than the expected accuracy of the flux density scale), finding one such strongly variable source in the \emerlin\ observations and two in the VLA observations, and modelled and subtracted these from the individual epochs (see \S\,\ref{sect:uvsub}). The flux densities of the remaining (non-variable) sources were then compared to assess for epoch-to-epoch errors on the global flux density scale. We found the individual epochs to be broadly consistent, with the average integrated flux densities of non-variable sources differing by less than $\sim 10$\%. Nevertheless, to correct these small variations, a gain table was generated and applied to bring each epoch to a common flux density scale (taken from the \emerlin\ epoch with the lowest rms noise, $\sigma_{\rm 1.5\,GHz}$).

In addition, the astrometry of each epoch was compared and aligned to the astrometric solutions derived by recent European VLBI Network (EVN) observations of the GOODS-N field \citep{radcliffe18}. By comparing the positions of 22 EVN-detected sources which are also in \emerge, we measured a systematic linear offset of $\sim 15$\,mas in RA (corresponding to $\sim 5\%$ of the $0\farcs3$ \emerlin\ PSF and $\sim 1\%$ of the $1\farcs5$ VLA PSF). This offset does not vary between epochs, and no correlation in the magnitude of the offset with the distance from the pointing centre was found, which indicates there are no significant stretch errors in the field. We determined that this offset arose due to an error in the recorded position of the phase reference source \citep{radcliffe18}, and corrected for this by applying a linear 15\,mas shift to the \emerlin\ datasets. In this manner, we have astrometrically tied the \emerge\ DR1 $uv$ data and images to the International Celestial Reference Frame (ICRF) to an accuracy of $\leqslant10$\,mas.

\begin{figure}
\centering
\includegraphics[width=\columnwidth]{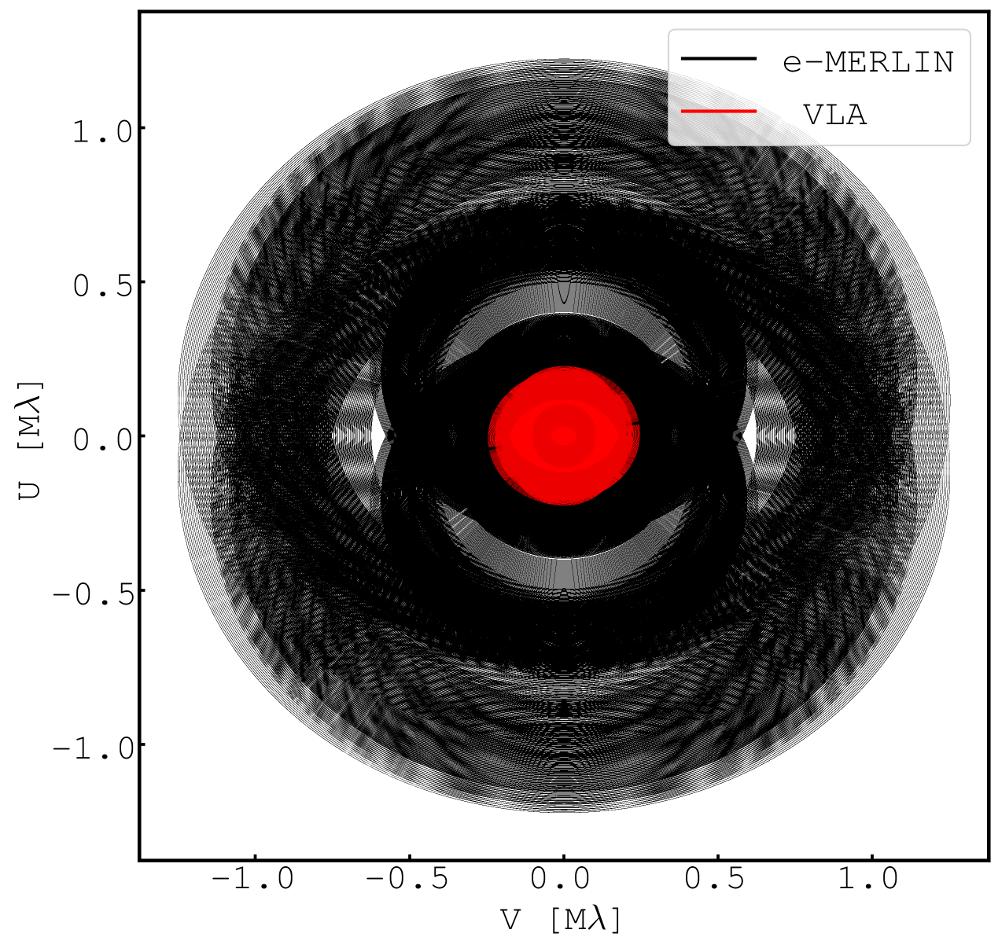}
\caption{$uv$ coverage of the combined \emerlin\ plus VLA 1.5\,GHz dataset presented in \S\,\ref{sect:data-all}. The long ($B_{\rm max}\sim 217$\,km) baselines of \emerlin\ hugely extend the VLA-only $uv$ coverage, while the presence of short baselines from the VLA ($B\sim0.68$--$36.4$\,km) overlap and fill the inner gaps in \emerlin's $uv$ coverage due to its shortest usable baseline length of $B_{\rm min}\sim10$\,km. The combined resolving power of both arrays provides seamless imaging capabilities with sensitivity to emission over $\sim0\farcs2$--$\,40\arcsec$ spatial scales.}\label{fig:uv_coverage}
\end{figure}

\begin{figure*}
\begin{centering}
\includegraphics[width=\textwidth]{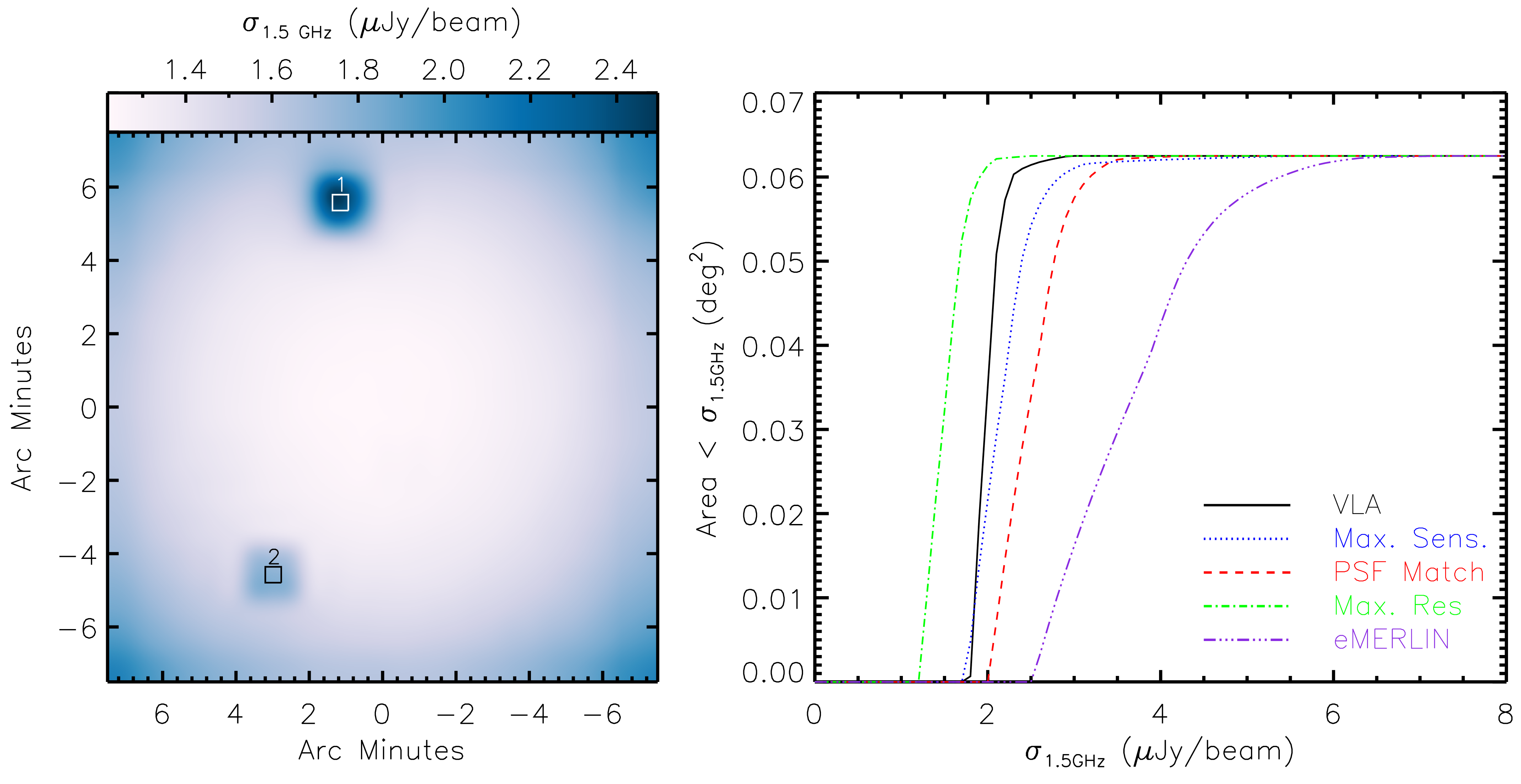}
\vspace*{-4mm}
\caption{\textit{Left}: Noise map ($\sigma_{\rm 1.5\,GHz}$) from our \emerge\ DR1 \emerlin+VLA naturally-weighted combination image (see Table\,\ref{tab:images}). Near the centre of the field our combination image reaches a noise level $\sigma_{\rm 1.5\,GHz}\sim 1.26\,\mu$Jy\,beam$^{-1}$, rising to $\sigma_{\rm 1.5\,GHz}\sim 2.1\,\mu$Jy\,beam$^{-1}$ at the corners of the field. The steady rise in $\sigma_{\rm 1.5\,GHz}$ with distance from the pointing centre reflects the primary beam correction applied to our combined-array images (see \S\,\ref{sect:primarybeam} for details). We note two regions of high noise within the \emerge\ DR1 analysis region, which surround the bright, \emerlin\ point sources J123659+621833 [1] and J123715+620823 [2] (the latter of which exhibits strong month-to-month variability). These elevated noise levels reflect the residual amplitude errors after our attempts to model and subtract these sources with {\sc uvsub} (see \S\,\ref{sect:uvsub} for details). \textit{Right}: Figure showing the total area covered in each \emerge\ DR1 1.5\,GHz image down to a given point source rms sensitivity, $\sigma_{\rm 1.5\,GHz}$. Note that point-source sensitivities are quoted in units ``per beam'', and therefore the naturally-weighted combined image (which has the smallest PSF of the images in this data release) has the lowest noise level per beam. The ``maximum sensitivity'' image has lower point source sensitivity but a larger beam, thereby giving it superior sentivity to emission on $\sim$arcsec scales. For \emerge\ DR1 our field of view is limited to the central $15\arcmin$\ of GOODS-N. In a forthcoming DR2, we aim to quadruple the survey area and double the sensitivity within the inner region.}
\label{fig:rms_examples}
\end{centering}
\end{figure*}

\subsection{VLA 1.5\,GHz}\label{sect:vla-reduction}

To both improve the point source sensitivity of our \emerge\ dataset and provide crucial short baselines needed to study emission on $\gtrsim 1\arcsec$ scales, 38\,hours (8 epochs of 4--6\,hours) of VLA L-Band data were obtained in 2011 Aug--Sep using the A-array configuration between $1\mbox{-}2\,{\rm GHz}$ (VLA project code TLOW0001). These data have been previously published by \citet{owen18}, and use a 1\,s integration time and 1\,MHz/channel frequency resolution, with 16\,spws of 64\,channels each, providing a total bandwidth of 1.024\,GHz. We retrieved the raw, unaveraged data from the archive and processed them using a combination of the VLA {\sc casa} pipeline \citep{mcmullin07}, along with additional manual processing steps. Initial flagging was performed using {\sc aoflagger} \citep{offringa2012}, before further automated flagging and initial calibration was applied using the VLA scripted pipeline packaged with {\sc casa} version 4.3.1. Flux density bootstrapping was performed using 3C\,286, while bandpass corrections were derived using the bright calibrator source 1313+6735 (which was also used for delay and phase tracking). After pipeline calibration the optimal data weights were derived based upon the rms scatter of the calibrated dataset. Finally, one round of phase-only self-calibration on each epoch of data was performed using a sky model of the central $5\arcmin$\ area (for which any resultant calibration errors due to the primary beam attenuation are expected to be minimal), and the data were exported with 3\,s time averaging.

The $uv$ coverage attained by combining these VLA observations with the \emerlin\ observations discussed in the previous section is shown in Fig.\,\ref{fig:uv_coverage}.

\subsection{Previous 1.4 GHz VLA + MERLIN observations}\label{sect:data-vla}

To maximise the sensitivity of the \emerge\ DR1 imaging products, we make use of earlier MERLIN and VLA $uv$ datasets obtained between 1996--2000, i.e.\ prior to the major upgrades carried out to both instruments in the last decade.  A total of 140\,hours of MERLIN and 42\,hours of pre-upgrade VLA (A-configuration) 1.5\,GHz data share the same phase centres as our more recent \emerlin\ and post-upgrade VLA observations. Full details of the data reduction strategies employed for these datasets are presented in \citet{Muxlow2005:hdf} and \citet{richards00}, respectively. These datasets have a much-reduced frequency coverage compared to the equivalent post-2010 datasets, i.e.\ the MERLIN observations have 0.5\,MHz/channel over 31 channels (yielding 15\,MHz total bandwidth) while the legacy VLA observations have 3.125\,MHz/channel over 14 channels (i.e.\ 44\,MHz total bandwidth).

These single-polarization legacy VLA and MERLIN datasets were not originally designed to be combined in the $uv$ plane, due to differences in channel arrangements of the VLA and MERLIN correlators. However, modern data processing techniques nevertheless allow this $uv$ plane combination to be achieved. We gridded both datasets onto a single channel (at a central reference frequency of 1.42\,GHz) by transforming the $u$, $v$ and $w$ coordinates from the multi-frequency synthesis gridded coordinates. This gridding ensures that the full $uv$ coverage is maintained during the conversion, with appropriate weights calculated in proportion to the sensitivity of each baseline within each array, and was performed within \aips\ by use of the \textsc{split} and \textsc{dbcon} tasks in a hierarchical manner. From these pseudo-single channel, single polarisation datasets, the data were then transformed into a Stokes I {\sc casa} Measurement Set format via the following steps: (i) A duplicate of each dataset was generated, with the designated polarisation converted from RR to LL; (ii) the \aips\ task \textsc{vbglu} was used to combine the two polarisations into one data set with two spws; (iii) the \aips\ task \textsc{fxpol} was used to re-assign the spws into a data set containing one spw with a single channel per polarisation. Finally, these data sets were then exported from \aips\ as {\sc uvfits} files and converted to Measurement Set format using the {\sc casa} task {\sc importuvfits}, to facilitate eventual $uv$ plane combination with the new \emerlin\ and VLA \emerge\ observations. We discuss the details of how our L-band data from both \emerlinmerlin\ and old/new VLA were combined in the $uv$ plane and imaged jointly in \S\,\ref{sect:imaging}.

\subsection{Subtraction of bright sources from 1.5\,GHz \emerlin\ and VLA data}\label{sect:uvsub}

The combination of extremely bright sources located away from the phase centre of an interferometer and small gain errors in the data (typically caused by primary beam attenuation and atmospheric variations across the field) can produce unstable sidelobe structure within the target field which cannot be deconvolved from the map, limiting the dynamic range of the final {\sc clean} map. These effects can be mitigated (while imaging) using direction-dependent calibration methods, such as \textsc{awprojection} \citep{bhatnagar2013}; however, without detailed models of the primary beam, this can be difficult (see \S\,\ref{sect:primarybeam} for a discussion of our current model of the \emerlin\ primary beam response). 

An alternative method of correcting these errors is to use an iterative self-calibration routine known as ``peeling'' \citep[e.g.][]{intema2009}, in which direction-dependent calibration parameters are determined and the source is modelled and subtracted from the visibility data. Initial exploratory imaging of our 1.5\,GHz VLA observations of GOODS-N revealed two bright sources ($S_{\rm 1.4\,GHz}\gtrsim 100$\,mJy; more than $10^5\times$ the representative rms noise at the centre of the field) which caused dynamic range problems of the kind described above. These sources -- J123452+620236 and J123538+621932 -- lie $7\arcmin$\ and $1\arcmin$\ outside the \emerge\ DR1 field, respectively. Due to the structures of these sources (i.e.\ unresolved by VLA and marginally-resolved by \emerlin), this issue disproportionately affected the VLA observations. To mitigate their effect on the target field, we adopted a variant of the peeling routine consisting of the following steps: (i) for each source and in each spectral window, an initial VLA-only model was generated (i.e.\ 32 model images covering 16\,spws for two sources); (ii) using these multi-frequency sky models, gain corrections were derived to correct the visibilities at the locations of the bright sources; (iii) the corrected bright sources were re-modelled. Because these sources lie outside the DR1 field, the Fourier transforms of these corrected models were then removed from the $uv$ data\footnote{By removing these sources from the $uv$ data we avoid the need to {\sc clean} them during deconvolution, significantly reducing the area to be imaged (and thus the computational burden) without loss of information on the target field.}. Finally, (iv) the gain corrections were inverted and re-applied to the visibilities such that the gains are again correct for the target field.

\begin{figure}
    \centering
    \includegraphics[width=\columnwidth]{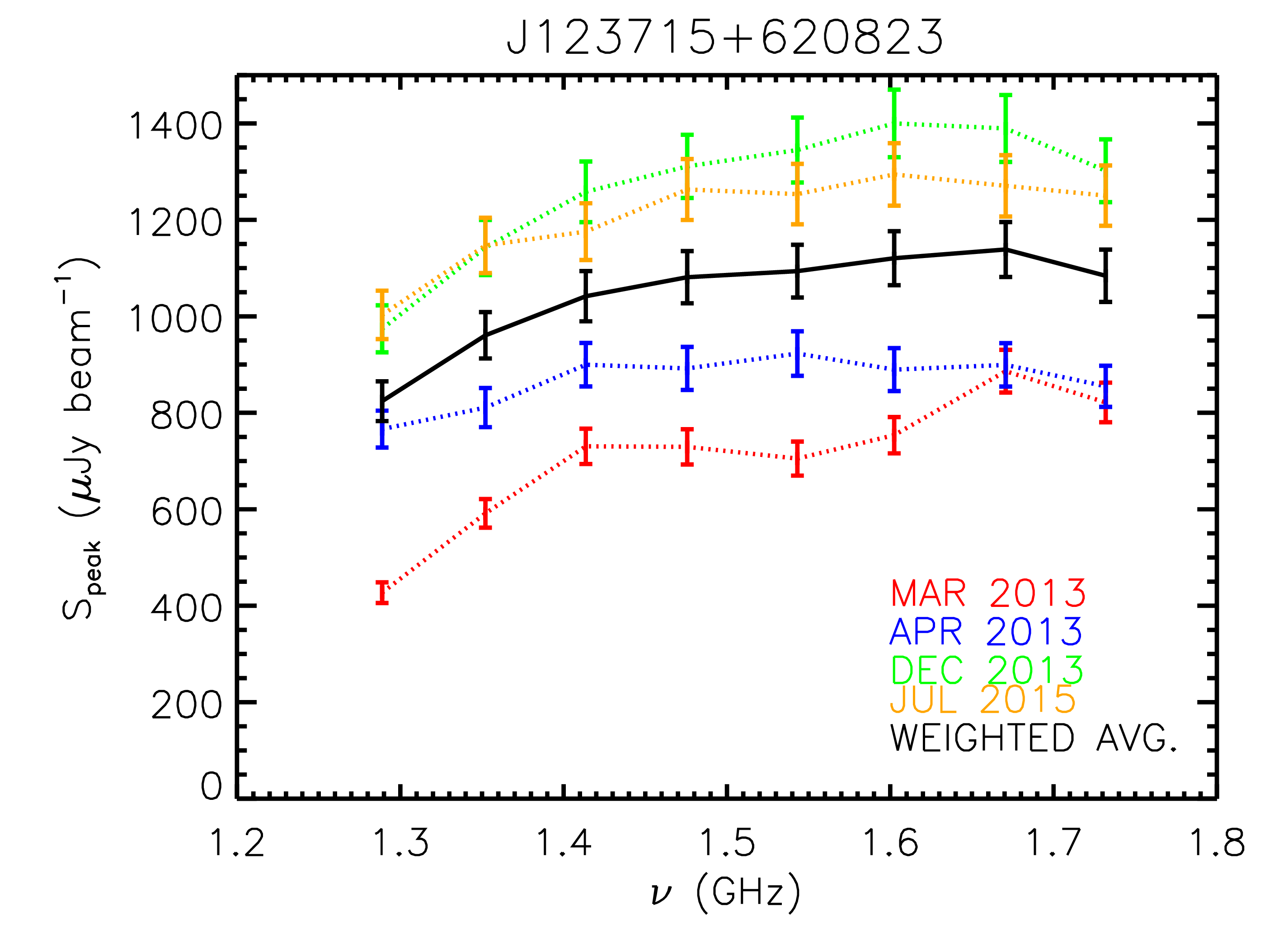}
    \vspace*{-5mm}
    \caption{Peak flux densities in eight frequency intervals across four epochs (Mar\,2013--Jul\,2015) for the \emerlin\ variable unresolved source J123715+620823. Due to small gain errors in the data it was necessary to iteratively self-calibrate (``peel'') this bright ($\sim 10^5\times$ the noise level at the centre of the field) point source epoch-by-epoch using a multi-frequency sky model. After peeling, we reinjected the source back in to our $uv$ data using the sensitivity-weighted average flux in each spectral window (solid black line).}
    \label{fig:variable}
\end{figure}

With these sources removed from the VLA data, further exploratory imaging of the 1.5\,GHz data revealed that two in-field sources (J123659+621833 and J123715+620823) caused significant image artefacts, but only in the \emerlin\ data (see Fig.\,\ref{fig:rms_examples}). We found the flux density of J123659+621833 to be constant (within $\leqslant10\%$) across all epochs with \emerlin\ observations (i.e.\ a two year baseline; see Table\,\ref{tab:e-MERGE-Data-products}), and so created one model for each of \emerlin's 8 spectral windows for this source, which we subtracted from the data following the procedure outlined above. On the other hand, image-plane fitting of J123715+620823 showed it to have both strong in-band spectral structure and significant short-term variability, increasing in peak flux density from $S_{\rm 1.5\,GHz}=730\pm36\,\mu$Jy to $S_{\rm 1.5\,GHz}=1311\pm26\,\mu$Jy across the nine months from Mar--Dec\,2013 before dropping to $S_{\rm 1.5\,GHz}=1249\pm63\,\mu$Jy by Jul\,2015 (see Fig.\,\ref{fig:variable}). J123715+620823 was also observed with the EVN during 5-6\,Jun 2014 by \citet{radcliffe18}, who measured a peak flux density $S_{\rm 1.5\,GHz}=2610\pm273\,\mu$Jy, thereby confirming the classification of J123715+620823 as a strongly-variable point source. To avoid amplitude errors in the model because of this strong source variability, it was necessary to create a model for J123715+620823 for each spectral window for each epoch of \emerlin\ data in order to derive gain corrections which are appropriate for that epoch. After subtracting the appropriate model of J123715+620823 from each epoch of \emerlin\ data, we then restored the source to the $uv$ data using a single flux-averaged model. This peeling process significantly reduced the magnitude and extent of the imaging artefacts around both J123659+621833 and J123715+620823, however some residual artefacts remain (Fig.\,\ref{fig:rms_examples}).

\subsection{Wide-field Integrated Imaging with \emerlin\ and VLA}\label{sect:imaging}

\begin{figure*}
\centering
\includegraphics[width=15.0cm]{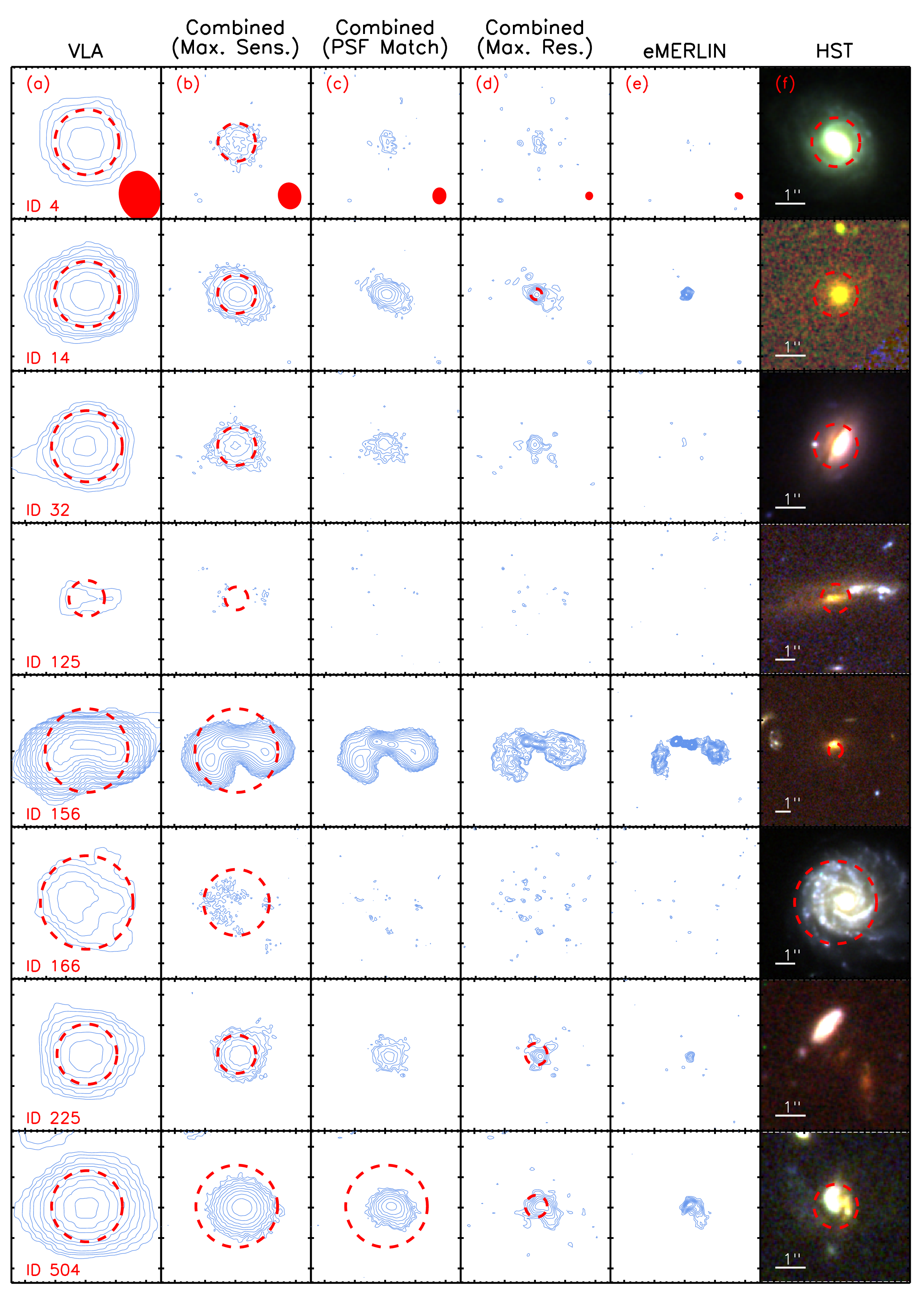}
\vspace*{-4mm}
\caption{Thumbnail images of 8 representative sources (one per row) from the \emerge\ DR1 catalogue of 848 radio sources (Thomson \etal, in prep), highlighting the need for a suite of radio images made with different weighting schemes (each offering a unique trade-off between angular resolution and sensitivity) to fully characterise the extragalactic radio source population. Columns (a)--(e) step through the five \emerge\ DR1 1.5\,GHz radio images in order of increasing angular resolution from VLA-only to \emerlin-only (see Table\,\ref{tab:images} for details). Contours begin at $3\sigma$ and ascend in steps of $3\sqrt{2}\times\sigma$\ thereafter, and the fitted Petrosian size (if statistically significant) is shown as a red dashed circle (see \S\,\ref{sect:sizes}). Column (f) shows three-colour (F606W, F814W, F850LP) \textit{HST} CANDELS thumbnail images for each source, with the optical Petrosian size shown as a red dashed circle. A 1\farcs0 scale bar is shown in white in each colour thumbnail. Together, columns (a)-(f) highlight the diversity of the \emerge\ DR1 source population, including a mixture of core-dominated AGN within quiescent host galaxies (ID\,14), merger-driven star-forming galaxies (ID\,125, 225), high-redshift wide-angle tail AGN (ID\,156) and face-on spiral galaxies (ID\,166).}\label{fig:eMERGE_5maps}
\end{figure*}

The primary goal of eMERGE is to obtain high surface brightness sensitivity ($\sigma_{\rm 1.5\,GHz}\lesssim 5\,\mu$Jy\,arcsec$^{-2}$) imaging at sub-arcsecond resolution across a field-of-view that is large enough ($\gtrsim 15\arcmin\times 15\arcmin$) to allow a representative study of the high-redshift radio source population. This combination of observing goals is beyond the capabilities of either \emerlin\ or VLA individually, hence the combination of data from VLA and \emerlin\ is essential. 

While co-addition of datasets obtained at different times from different array configurations of the same telescope (e.g.\ VLA, ALMA, ATCA) is a routine operation in modern interferometry, the differing internal frequency/polarisation structures of our new \emerlin/VLA and previously-published MERLIN/VLA datasets prohibited a straightforward concatenation of the datasets using standard (e.g.\ \aips\ {\sc dbcon} or {\sc casa} {\sc concat}) tasks.

Historically, circumventing this issue has necessitated either image-plane combination of datasets, or further re-mapping of the internal structures of the $uv$ datasets to allow them to be merged.

The former approach involves generating dirty maps (i.e.\ with no {\sc clean}ing/deconvolution) from each dataset independently, co-adding them in the image plane, and then deconvolving the co-added map using the weighted average of the individual PSFs using the \citet{hogbom74} {\sc clean} algorithm, as implemented in the \aips\ task {\sc apcln} \citep[e.g.\ ][]{Muxlow2005:hdf}. While this approach sidesteps difficulties in combining inhomogeneous datasets properly in the $uv$ plane -- and produces reliable results for sources whose angular sizes are in the range of scales to which both arrays are sensitive ($\theta\sim 1$--$1\farcs5$) -- the fidelity of the resulting image is subject to the reliance on purely image-based deconvolution using ``minor cycles'' only. This is a potentially serious limitation when imaging structures for which only one array provides useful spatial information (i.e.\ extended sources which are resolved-out by \emerlin\ or compact sources which are unresolved by VLA), where {\sc clean}ing using a hybrid beam is not the appropriate thing to do.

An alternative approach -- used by \citet{biggs08} -- is to collapse the multi-frequency datasets from each telescope along the frequency axis, preserving the $uvw$ coordinates of each visibility (as was done for the pre-2010 VLA data described in \S\,\ref{sect:data-vla}), and then concatenate and image these single-channel datasets. This approach allows the $uv$ coverage of multiple datasets to be combined, bypassing the issues with image-plane combination and allowing a single imaging run to be performed utilising the \citet{schwab84} {\sc clean} algorithm (i.e.\ consisting of both major and minor {\sc clean} cycles). However, while this approach has proved successful when combining together MERLIN/VLA datasets of relatively modest bandwidth, the technique of collapsing the available bandwidth down to a single frequency channel implicitly assumes that the source spectral index is flat across the observed bandwidth. While this condition is approximately satisfied for most sources given the narrow bandwidths of the older MERLIN/VLA datasets, it cannot be assumed given the orders-of-magnitude increase in bandwidth which is now available with both instruments. For sources with non-flat spectral indexes, this approach would introduce amplitude errors in the final image.

In order to successfully merge our \emerlinmerlin\ and old/new VLA datasets we use {\sc wsclean} \citep{Offringa2014:ws}, a fast, wide-field imager developed for imaging data from modern synthesis arrays. {\sc wsclean} utilises the $w$-stacking algorithm, which captures sky curvature over the wide field of view of \emerlin\ by modelling the radio sky in three dimensions, discretising the data along a vector $w$ (which points along the line of sight of the array to the phase centre of the observations), performs a Fourier Transform on each $w$-layer and finally recombines the $w$-layers in the image plane. In addition to offering significant performance advantages over the {\sc casa} implementation of the $w$-projection algorithm \citep[for details, see][]{Offringa2014:ws}, {\sc wsclean} also possesses the ability to read in multiple calibrated Measurement Sets from multiple arrays (with arbitrary frequency/polarisation setups) and grid them on-the-fly, sidestepping the difficulties we encountered when trying to merge these datasets using standard \aips/{\sc casa} tasks. {\sc wsclean} allows us to generate deep, wide-field images using all the 1--2\,GHz data from both arrays (spanning a 20\,year observing campaign) in a single, deep imaging run, deconvolving the resulting (deterministic) PSF from the image using both major and minor cycles, and without loss of frequency or polarisation information.

\subsubsection{Data weights}

The \emerge\ survey was conceived with the aim that -- upon completion -- the naturally-weighted combined-array 1.5\,GHz image would yield a PSF that could be well-characterised by a 2-dimensional Gaussian function, with minimal sidelobe structure. In this survey description paper for Data Release 1, we present imaging which utilises $\sim 90$\% of the anticipated VLA 1.5\,GHz data volume, but with only $\sim 25$\% of the \emerlin\ observations included. As a result, the PSF arising from our naturally-weighted combined dataset more closely resembles the superposition of two 2-dimensional Gaussian components -- one, a narrow ($\theta_{\rm res}\sim 0\farcs2$) component representing the \emerlin\ PSF, and the other, a broader ($\theta_{\rm res}\sim 1\farcs5$) component representing the VLA A-array PSF -- joined together with significant shoulders at around $\sim 50$\% of the peak\footnote{It is expected that the inclusion of $\sim 4\times$ more \emerlin\ data in \emerge\ DR2 will smooth out the shoulders of the naturally-weighted combined PSF and achieve our long-term goal of a Gaussian PSF.}.  

Standard {\sc clean} techniques to deconvolve the PSF from an interferometer dirty image entail iteratively subtracting a scaled version of the true PSF (the so-called ``dirty beam'') at the locations of peaks in the image, building a model of delta functions (known as ``clean components''), Fourier transforming these into the $uv$ plane and subtracting these from the data. This process is typically repeated until the residual image is noise-like, before the clean components are restored to the residual image by convolving them with an idealised (2-dimensional Gaussian) representation of the PSF. The flux density scale of the image is in units of Jy\,beam$^{-1}$, where the denominator is derived from the volume of the fitted PSF. While this approach works well for images where the dirty beam closely resembles a 2-dimensional Gaussian to begin with, great care must be taken if the dirty beam has prominent shoulders. In creating our {\sc clean}ed naturally-weighted \emerlinmerlin\ plus VLA combination image, we subtracted scaled versions of the true PSF at the locations of positive flux and then restored these with an idealised Gaussian, whose fit is dominated by the narrow central portion of the beam produced by the \emerlin\ baselines. The nominal angular resolution of this naturally-weighted combination image is $\theta_{\rm res}=0\farcs28\times 0\farcs26$, with a beam position angle of $84$\,deg, and the image has a representative noise level of $\sigma_{\rm 1.5\,GHz}=1.17\,\mu$Jy\,beam$^{-1}$. However, subsequent flux density recovery tests comparing the VLA-only and the \emerlin+VLA combination images revealed that while this process works well for bright, compact sources, (recovering $\sim 100$\% of the VLA flux density but with $\sim 5\times$ higher angular resolution), our ability to recover the flux density in fainter, more extended ($\gtrsim 0\farcs7$) sources is severely compromised. This is because the representative angular resolutions of the clean component map and the residual image (on to which the restored clean components are inserted) are essentially decoupled (due to the restoring beam being a poor fit to the ``true'', shouldered PSF). As a result of this, faint radio sources restored at high resolution are imprinted on $\sim$ arcsecond noise pedestals, containing the residual un-{\sc clean}ed flux density in the map. This limits the ability of source-fitting codes to find the edges of faint radio sources in the naturally weighted image, with a tendency to artificially boost their size and flux density estimates. Moreover, the difference in the effective angular resolutions of the clean component and residual images renders the map units themselves (Jy\,beam$^{-1}$) problematic. This issue will be discussed in more detail in the forthcoming \emerge\ catalogue paper (Thomson \etal, in prep), however we stress that in principle it applies to \textit{any} interferometer image whose dirty beam deviates significantly from a 2-dimensional Gaussian.

To mitigate this effect, a further two 1.5\,GHz combined-array images were created with the aim of smoothing out the shoulders of the naturally-weighted \emerlin+VLA PSF. We achieved this by using the {\sc wsclean} implementation of ``Tukey'' tapers \citep{tukey62}. Tukey tapers are used to adjust the relative contributions of short and long baselines in the gridded dataset, and work in concert with the more familiar \citet{briggs95} {\sc robust} weighting schemes. They can be used to smooth the inner or outer portions of the $uv$ plane (in units of $\lambda$) with a tapered cosine window which runs smoothly from 0 to 1 between user-specified start ({\rm UVm}) and end points ({\rm iTT})\footnote{see \url{https://sourceforge.net/p/wsclean/wiki/Tapering/} for details}. By effectively down-weighting data on certain baselines, the output image then allows a different trade-off between angular resolution, rms sensitivity per beam, and dirty beam Gaussianity to be achieved.

To provide optimally sensitive imaging of extended $\mu$Jy radio sources while retaining $\sim$kpc-scale (i.e.\ sub-arcsecond) resolution, we complement the naturally-weighted \emerlin+VLA combination image with two images which utilise Tukey tapers:

\begin{enumerate}
    \item We create a maximally-sensitive combination image using both inner and outer Tukey tapers (${\rm UVm}=0\lambda$ and ${\rm iTT}=82240\lambda$) along with a {\sc briggs} robust value of 1.5. The angular resolution of this image is $\theta_{\rm res}=0\farcs89\times 0\farcs78$ at a position angle of 105\,deg and with an rms sensitivity $\sigma_{\rm 1.5\,GHz}=1.71\,\mu$Jy\,beam$^{-1}$ (corresponding to $\sim 2.46\,\mu$Jy\,arcsec$^{-2}$).
    \item To exploit the synergy between our 1.5\,GHz and 5.5\,GHz datasets (and thus to enable spatially-resolved spectral index work), we have identified a weighting scheme which delivers a 1.5\,GHz PSF that is close to that of the VLA 5.5\,GHz mosaic image of \citet{Guidetti2017:em}. We find that a combination of a Briggs taper with {\sc robust}$=1.5$ and a Tukey taper with ${\rm UVm}=0\lambda$, ${\rm iTT}=164480\lambda$ yields a two-dimensional Gaussian PSF of size $\theta_{\rm res}=0\farcs55\times 0\farcs42$ at a position angle of 112\,deg. To provide an exact match for the 5.5\,GHz PSF ($\theta_{\rm res}=0\farcs56\times 0\farcs47$ at a position angle of 88\,deg) we use this weighting scheme in combination with the --{\sc beam}-{\sc shape} parameter of {\sc wsclean}. The resulting rms of this image is $\sigma_{\rm 1.5\,GHz}=1.94\,\mu$Jy\,beam$^{-1}$, or $\sim 7.37\,\mu$Jy\,arcsec$^{-2}$.
\end{enumerate}

Together with VLA-only and \emerlin-only images (representing the extremes of the trade-off in sensitivity and resolution), these constitute a suite of five 1.5\,GHz images that are optimised for a range of high-redshift science applications (see Table\,\ref{tab:images}).

The trade-off in angular resolution versus sensitivity between these five weighting schemes is highlighted for a representative subset of \emerge\ sources in Fig.\,\ref{fig:eMERGE_5maps}. 

\subsubsection{Primary beam corrections for combined-array images}\label{sect:primarybeam}

The primary beam response of a radio antenna defines the usable field of view of a single-pointing image made with that antenna. In the direction of the pointing centre, the primary beam response is unity, dropping to $\sim 50$\% at the half power beam width ($\theta_{\rm HPBW}\sim \lambda/D$ for an antenna diameter $D$). For wide-field images it is essential to correct the observed flux densities of sources observed off-axis from the pointing centre for this primary beam response.

In the case of homogeneous arrays (such as VLA), the primary beam response of the array is equivalent to that of an individual antenna. Moreover, because the antennas are identical, the primary beam response of the array is invariant to the fraction of data flagged on each antenna/baseline. Detailed primary beam models for the VLA in each antenna/frequency configuration are incorporated in {\sc casa} and can be implemented on-the-fly during imaging runs by setting {\sc pbcor}=True in {\sc tclean}, or can be exported as {\sc fits} images using the {\sc widebandpbcor} task. However for inhomogeneous arrays (such as \emerlin) the primary beam response is a sensitivity-weighted combination of the primary beam responses from each antenna pair in the array. These weights are influenced by the proportion of data flagged on each antenna/baseline, and thus vary from observation to observation.

To correct our \emerlin\ observations for the primary beam response, we constructed a theoretical primary beam model based on the weighted combination of the primary beams for each pair of antennas in the array. This model is presented in detail in \citet{wrigley16} and Wrigley \etal\ (in prep), however we provide an outline of our approach here. To model the primary beam of \emerlin, we first derived theoretical 2-dimensional complex voltage patterns $V_i V^{\star}_j$ and $V^{\star}_i V_j$ for each pair of antennas $ij$ based on knowledge of the construction of the antennas (effective antenna diameters, feed blocking diameters, illumination tapers, pylon obstructions and spherical shadow projections due to the support structures for the secondary reflector). We checked the fidelity of these theoretical voltage patterns via holographic scans, wherein each pair of antennas in the array was pointed in turn at a bright point source (e.g.\ 3C\,84), with one antenna tracking the source while the other scanned across it in a raster-like manner, nodding in elevation and azimuth to map out the expected main beam.

Next, we extracted the mean relative baseline weights $\langle\sigma_{ij}\rangle$ for each pair of antennas $ij$ recorded in the Measurement Set (post-flagging and post-calibration), and constructed the power beam $P_{ij}$ for each antenna pair from these complex voltage patterns $V_i$, $V_j$:

\begin{equation}\label{eq:power-beam}
P_{ij} = \frac{[V_{i}V_{j}^{*} + V_{i}^{*}V_{j}]}{2\sqrt{\langle\sigma_{ij}\rangle}}
\end{equation}

Finally, the primary beam model for the whole array, $P_B$, was constructed by averaging each baseline beam around the axis of rotation (simulating a full 24\,hour \emerlin\ observing run) and summing each of these weighted power beam pairs:

\begin{center}
    \begin{equation}
        P_B = \sum\limits_{i<j} P_{ij}
    \end{equation}
\end{center}

This primary beam model comprises a 2-dimensional array representing the relative sensitivity of our \emerlin\ observations as a function of position from the pointing centre; the model is normalised to unity at the pointing centre, and tapers to $\sim 57$\% at the corners of our DR1 images, a distance of $\sim 11\arcmin$ from the pointing centre. We applied this primary beam correction to the images made using {\sc wsclean} in the image plane, dividing the uncorrected map by the beam model.

To construct an appropriate primary beam model for our \emerlin+VLA combination images, we exported the 2-dimensional VLA primary beam model from {\sc casa}, re-gridded it to the same pixel scale as our \emerlin\ beam model and then created sensitivity-weighted combinations of the \emerlin+VLA primary beam for each of the DR1 images listed in Table\,\ref{tab:e-MERGE-Data-products}. We again applied these corrections by dividing the {\sc wsclean} combined-array maps by the appropriate primary beam model. 

The effect of applying these primary beam models is an elevation in the noise level (and in source flux densities) in the corrected images as a function of distance from the pointing centre, which is highlighted in Fig.\,\ref{fig:rms_examples}.

\begin{table*}
\begin{centering}
\caption[\emerge\ DR1 image summary]{\emerge\ DR1 image summary}
\label{tab:images}
\begin{tabular}{lccccccccccc}
\hline
\multicolumn{1}{l}{Image name} &
\multicolumn{1}{c}{Description} &
\multicolumn{1}{c}{Frequency} & 
\multicolumn{1}{c}{Synthesized beam} &
\multicolumn{1}{c}{$\sigma_{\rm rms}^a$} & \\
\multicolumn{1}{l}{} &
\multicolumn{1}{c}{} &
\multicolumn{1}{c}{} &
\multicolumn{1}{c}{} &
\multicolumn{1}{c}{($\mu$Jy\,beam$^{-1}$)} &\\
\hline
VLA & Naturally-weighted & 1.5\,GHz & $1\farcs68\times1\farcs48$ @ 105.88$^\circ$ & $2.04$ \\
Combined (Max. Sens.) & \emerlin +VLA combined-array image, & '' & $0\farcs89\times0\farcs78$ @ 105$^\circ$ & $1.71$\\
 & weighted for improved sensitivity & & & \\
Combined (PSF Match) & \emerlin +VLA, weighted to match VLA & '' & $0\farcs56\times0\farcs47$ @ 88$^\circ$ & $1.94$\\
 & 5.5\,GHz resolution for spectral index work & & & \\
Combined (Max. Res.) & \emerlin +VLA, weighted for & '' & $0\farcs28\times0\farcs26$ @ 84$^\circ$ & $1.17$\\
 & improved angular resolution & & & \\
\emerlin & \emerlin-only, naturally-weighted & '' & $0\farcs31\times 0\farcs21$ @ 149$^\circ$ & $2.50$ \\
\hline
C-band mosaic & 5.5\,GHz, naturally-weighted mosaic$^b$ & 5.5\,GHz & $0\farcs56\times 0\farcs47$ @ 88$^\circ$ & $1.84$ \\
\hline
\end{tabular}
\end{centering}
\\ {\small Notes: $^a\sigma_{\rm rms}$ values are in units of $\mu$Jy\,beam$^{-1}$, and are therefore dependent on the beam size -- the ``max res'' combination image has the lowest $\sigma_{\rm rms}$ (and therefore, the best point-source sensitivity of all images in this Data Release), however the small beam limits its sensitivity to extended emission, to which the lower-resolution combined-array images -- with slightly higher $\sigma_{\rm rms}$ -- are more sensitive. $^b$Previously published by \citet{Guidetti2017:em}. }
\end{table*}

\subsubsection{Time and bandwidth smearing}\label{sect:smearing}

As discussed in \ref{sect:emerlin-dr}, the quantisation of astrophysical emission by an interferometer into discrete time intervals and frequency channels results in imprecisions in the $(u,v)$ coordinates of the recorded data with respect to their true values. Both time and frequency quantisation have the effect of distorting the synthesized image in ways that cannot be deconvolved analytically using a single, spatially-invariant deconvolution kernel. The effect is a  ``smearing'' of sources in the image plane, which conserves their total flux densities but lowers their peak flux densities. Time/bandwidth smearing are an inescapable aspect of creating images from \textit{any} interferometer, but the effects are most significant in wide-field images, particularly on longer baselines and for sources located far from the pointing centre \citep[e.g.\ ][]{bridle1999:in}.

In order to compress the data volume of \emerge\ and ease the computational burden of imaging, we averaged our \emerlin\ observations by a factor $4\times$\ (from a native resolution of 0.125\,MHz/channel to 0.5\,MHz/channel), but did not average the data in time beyond the 1\,s/integration limit of the \emerlin\ correlator. We did not average the VLA observations in frequency beyond the native 1\,MHz/channel resolution, but did average in time to 3\,s/integration (as described in \S\,\ref{sect:vla-reduction}).

Using the SimuCLASS interferometry simulation pipeline developed by \citet{harrison20} we empirically determine that on the longest \emerlin\ baselines, at a distance of $10\farcm6$ from the pointing centre, bandwidth smearing induces a drop in the peak flux density of a point source of up to $\sim20\%$. This result - which is in agreement with the analytical relations in \citet{bridle1999:in} - limits the usable field-of-view of these data to the $15\arcmin\times15\arcmin$\ region overlying the \textit{HST} CANDELS region of GOODS-N.  By including the shorter baselines of the VLA, this smearing is reduced significantly, to: (i) $\sim 4$\% in the VLA-only image\footnote{in agreement with the performance specification of the VLA: \url{https://science.nrao.edu/facilities/vla/docs/manuals/oss/performance/fov}}; and (ii) $\lesssim 8\%$ at the edges of the ``maximum sensitivity'' DR1 combination image.

The frequency averaging of our \emerlin\ observations -- which was necessary in order to image the data using current compute hardware -- is therefore the primary factor limiting the usable \emerge\ DR1 field of view to that of the 76\,m Lovell Telescope. We note that in order to fully image the \emerlin\ observations out to the primary beam of the 25\,m antennas (as is planned for \emerge\ DR2) it will be necessary to re-reduce these data with no frequency averaging applied.

\subsubsection{VLA 5.5\,GHz}
Included in the \emerge\ DR1 release is the seven-pointing VLA 5.5\,GHz mosaic image of GOODS-N centred on J2000 RA $12^{h}36^{m}49\fs4$ DEC $+62^{\circ}12^{\prime}58\farcs0$, which was previously published by \citet[][in which a detailed description of the data reduction and imaging strategies is presented]{Guidetti2017:em}. For completeness, these observations are briefly summarised below.

The GOODS-N field was observed at 5.5\,GHz with the VLA in the A- and B-configuration, for 14\,hrs and 2.5\,hrs respectively. The total bandwidth of these observations is 2\,GHz, comprised of 16 spws of 64 channels each (corresponding to a frequency resolution of 2\,MHz/channel).

These data were reduced using standard \aips\ techniques, with the bright source J1241+6020 serving as the phase reference source and with 3C\,286 and J1407+2828 (OQ\,208) as flux density and bandpass calibrators respectively. Each pointing was imaged separately using the {\sc casa} task {\sc tclean}, using the multi-term, multi-frequency synthesis mode ({\sc mtmfs}) to account for the frequency dependence of the sky model. These images were corrected for primary beam attenuation using the task \textsc{widebandpbcor} and then combined in the image plane to create the final mosaic using the \aips\ task {\sc hgeom}, with each pointing contributing to the overlapping regions in proportion to the local noise level of the individual images. The final mosaic covers a $13\farcm5$ diameter area with central rms of $\sigma_{\rm 5.5\,GHz}\lesssim 2\mu$\,Jy\,beam$^{-1}$, and has a synthesized beam $\theta_{\rm res}=0\farcs56\times 0\farcs47$ at a position angle of 88\,deg. 

A total of 94 AGN and star-forming galaxies were extracted above $5\sigma$, of which 56 are classified as spatially extended \citep[see ][ for details]{Guidetti2017:em}.

\subsection{Ancillary data products}
\subsubsection{VLA 10\,GHz}
To provide additional high-frequency radio coverage of a subset of the \emerge\ DR1 sources, we also use observations taken at 10\,GHz as part of the GOODS-N Jansky VLA Pilot Survey \citep{murphy17}. These observations (conducted under the VLA project code 14B-037) comprise a single deep pointing (24.5\,hr on source) towards $\alpha=12^{\rm h}36^{\rm m}51\fs21$, $\delta=+62^{\circ}13^{\prime}37\farcs4$, with approximately 23\,hours of observations carried out with the VLA in A-array and 1.5\,hours in C-array. We retrieved these data from the VLA archive and, following \citet{murphy17}, calibrated them using the VLA {\sc casa} pipeline (included with {\sc casa} $v$ $4.5.1$). 3C\,286 served as the flux and bandpass calibrator source and J1302+5728 was used as the complex gain calibrator source. 

We created an image from the reduced $uv$ data with {\sc wsclean} using natural weighting, which includes an optimised version of the multiscale deconvolution algorithm \citep{cornwell08, offringa17} to facilitate deconvolution of the VLA beam from spatially-extended structures. Our final image covers the VLA X-band primary beam ($6\arcmin$\ in diameter) down to a median rms sensitivity of $\sigma_{\rm 10\,GHz}=1.28\,\mu$Jy\,beam$^{-1}$ across the field (reaching $\sigma_{\rm 10\,GHz}=0.56\,\mu$Jy\,beam$^{-1}$ within the inner $0\farcm8\times0\farcm8$) and with a restoring beam that is well-approximated by a two-dimensional Gaussian of size $0\farcs27\times 0\farcs23$ at a position angle of 4\,deg.

\subsubsection{Optical--near-IR observations}

In order to derive key physical properties (e.g.\ photometric/spectroscopic redshift information and stellar masses) of the host galaxies associated with the \emerge\ DR1 sample we utilise the rich, multi-wavelength catalogue of the GOODS-N field compiled by the 3D-HST team \citep{brammer12, Skelton2014;3DHST}. This includes seven-band \textit{Hubble} Space Telescope (\textit{HST}) imaging from the 3D-HST, Cosmic Assembly Near-infrared Deep Extragalactic Legacy Survey \citep[CANDELS: ][]{grogin11,koekemoer11} and GOODS \citep{giavalisco14} projects, along with a compilation of ancillary data from the literature including: (i) Subaru Suprime-Cam $B, V, R_{c}, I_{c}, z^{\prime}$ and Kitt Peak National Observatory 4\,m telescope U-band imaging from the Hawaii-HDFN project \citep{capak04}; (ii) Subaru MOIRCS $J, H, K$ imaging from the MODS project \citep{kajisawa11}, and (iii) \textit{Spitzer} IRAC $3.6, 4.5, 5.8, 8.0\,\mu$m imaging from the GOODS and SEDS projects \citep{dickinson03,ashby13}.

We defer a detailed discussion of the multi-wavelength properties of the \emerge\ sample to future papers, but emphasise that the 3D-HST catalogue is used to provide photometric redshift information for the \emerge\ sample in the following sections. 

\section{Analysis, Results \& Discussion}\label{sect:results}

The detailed properties (and construction) of the \emerge\ DR1 1.5\,GHz source catalogue will be presented in detail in a forthcoming publication (Thomson \etal, in prep), however we present an overview of the catalogue properties here, including the 1.5\,GHz angular size measurements of $\sim 500$ star-forming galaxies and AGN at $z\gtrsim 1$. 

\subsection{Radio source catalogue}\label{sect:catalogue}

For the purposes of this survey description paper, we use the VLA 1.5\,GHz image to identify sources, as this image has the optimal surface brightness sensitivity to detect sources which are extended on the scales expected of high-redshift galaxies ($\gtrsim 0.5''$); we then measure the sizes and integrated flux densities of these VLA-identified sources in the higher-resolution 1.5\,GHz maps.

We extract source components from the VLA image using the {\sc pybdsf} package \citep{mohan15}, which (i) creates background and noise images from the data via boxcar smoothing, (ii) identifies ``islands'' of emission whose peaks are above a given signal-to-noise threshold, and (iii) creates a sky model by fitting a series of connected Gaussian components to each island in order to minimise the residuals with respect to the background noise. We identify the optimum signal-to-noise (S/N) threshold at which to perform source extraction following the procedure outlined by \citet{stach19}. Briefly, we create an ``inverted'' copy of the VLA 1.5\,GHz image by multiplying the original pixel data by -1, and perform {\sc pybdsf} source extraction runs on the real and inverted maps with S/N thresholds between $3$--$10\sigma$ (in steps of $0.2\times\sigma$). At each step, we record the number of detected sources in the real (i.e.\ \textit{positive}) map, $N_{P}$, as well as the number of sources detected in the inverted (i.e. \textit{negative}) map, $N_{N}$. By definition any source detected in the inverted image is a false-positive. To quantify the false-positive rate as a function of S/N, we measure the ``Purity'' parameter for each source-extraction run:

\begin{center}
  \begin{equation}
P = \frac{N_{P}-N_{N}}{N_{P}}
  \end{equation}
\end{center}

We find that the source catalogue has a Purity of 0.993 (i.e.\ a false-positive rate $\leq 1\%$) at a source detection threshold of $4.8\sigma$.

After visually inspecting the data, best-fit model and residual thumbnails for each extracted source, we found evidence that some sources exhibited significant residual emission which was not well fit, indicating that the morphologies of some sources are too complicated (even in the 1\farcs5 resolution VLA image) to be adequately modelled with Gaussian components alone. To improve the model accuracy, we re-ran the source extraction procedure with the {\sc atrous\_do} module enabled within {\sc pybdsf}. This module decomposes the residual image left after multi-component Gaussian fitting in to wavelet images in order to identify extended emission -- essentially ``mopping up'' the extended flux from morphologically complex sources -- and was used to produce the final VLA 1.5\,GHz flux density measurements for our \emerge\ DR1 source catalogue.

\subsection{Illustrative analysis of a representative high-redshift \emerge\ source}

\begin{figure*}
\begin{centering}
\includegraphics[width=\textwidth]{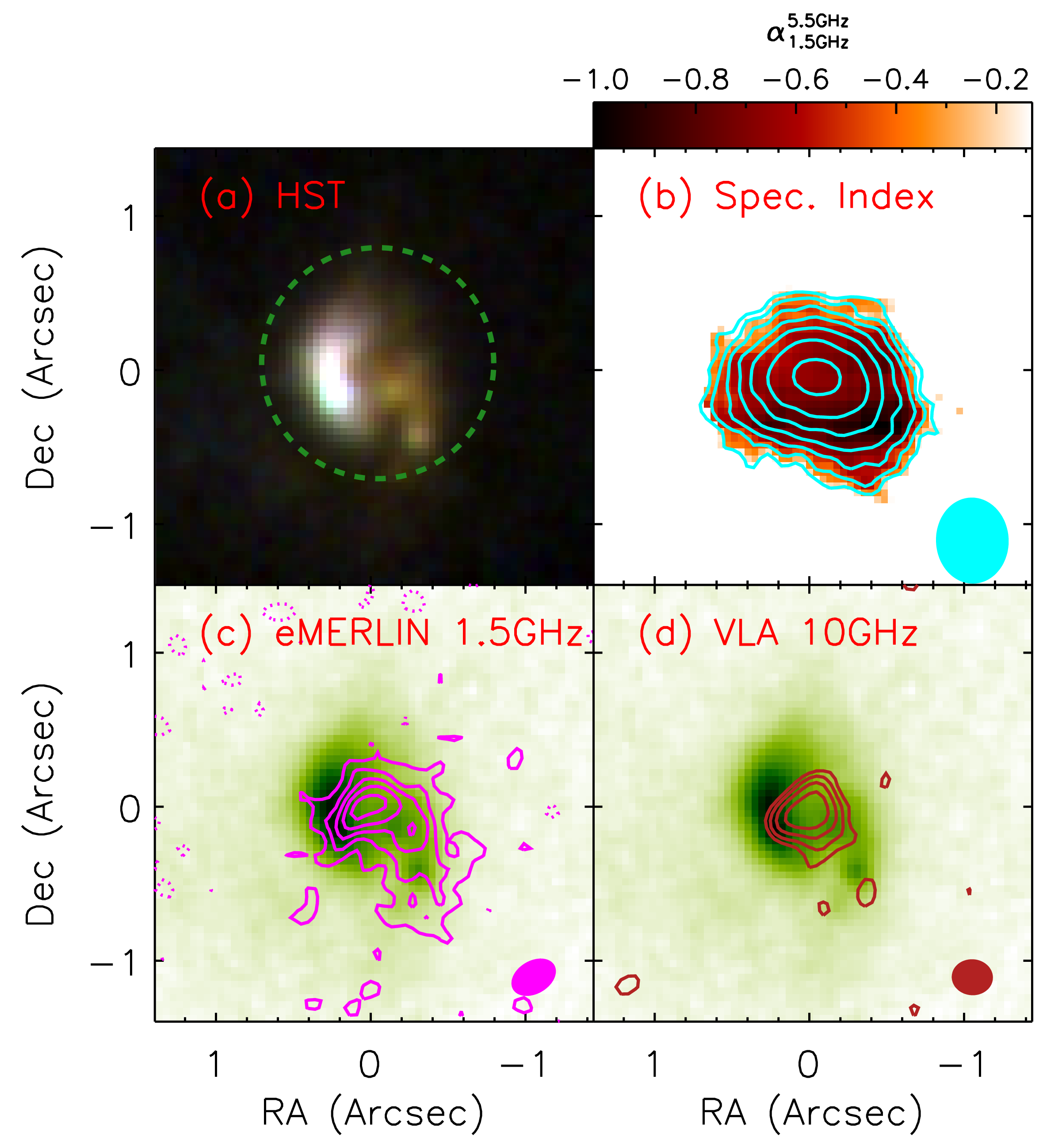}
\vspace*{-4mm}
\caption{Thumbnail images of the interacting system J123634+621241, dubbed the ``Seahorse'' galaxy. (\textit{a}): \textit{HST} three-colour (F606W, F814W, F850LP) image highlighting the disturbed morphology of this apparent close pair of merging galaxies. The green dotted circle has a radius of $1\farcs5$, representing the VLA 1.5\,GHz PSF. (\textit{b}): 1.5-to-5.5\,GHz spectral index ($\alpha^{\rm 5.5\,GHz}_{\rm 1.5\,GHz}$) image for the Seahorse (red heatmap) with cyan contours beginning at $3\times\sigma$ (and in steps of $\sqrt{2}\times\sigma$ thereafter for a local $\sigma=2.9\,\mu$Jy\,beam$^{-1}$) showing the 1.5\,GHz morphology in the PSF-matched \emerlin+VLA combination image (i.e.\ the 1.5\,GHz image with the same beam as the 5.5\,GHz VLA-only mosaic image, and which is used to create the spectral index image). The spectral index, $\alpha^{\rm 5.5\,GHz}_{\rm 1.5\,GHz}$ ranges between $-1.0 < \alpha^{\rm 5.5\,GHz}_{\rm 1.5\,GHz}<-0.1$. We see evidence that the redder optical galaxy is associated with steep spectrum ($\alpha<-1.0$) aged synchrotron emission in the radio tail, whilst the bluer optical galaxy is coincident with younger, less-steep ($\alpha\sim-0.7$) radio emission found in the bright extended nuclear starburst. The $0\farcs56 \times 0\farcs47$\ PSF is shown in the bottom-right corner with a cyan ellipse. (\textit{c}): \emerlin-only 1.5\,GHz contours of the Seahorse galaxy, plotted in magenta at $[-1,1,2,3,4,6...12]\times5.925\,\mu$Jy\,beam$^{-1}$ (i.e. $2.5\times\sigma_{\rm 1.5\,GHz}$) over the monochrome \textit{HST} F814W optical image. The peak of the radio emission likely traces the optically obscured nuclear starburst which is responsible for producing the far-IR emission in this system. The $0\farcs31\times 0\farcs21$ \emerlin\ PSF is shown. (\textit{d}): VLA 10\,GHz contours of the Seahorse galaxy \citep{murphy17} plotted in red over the monochrome \textit{HST} F814W image. The $0\farcs27\times0\farcs23$ PSF is shown. Contours begin at $3\times\sigma$ and in steps of $\sqrt{2}\times\sigma$\ thereafter for $\sigma=1.19\,\mu$Jy\,beam$^{-1}$. At these higher radio frequencies, there is very little extended emission visible from the evolved stellar population and instead we see redshifted free-free emission which directly traces the current active starburst.}
\label{fig:ant_thumbs}
\end{centering}
\end{figure*}

To highlight the science capabilities of our high angular resolution (sub-arcsecond) \emerge\ DR1 dataset, we present a short, single-object study of a representative source from our full catalogue of 848 sources. J123634+621241 (ID\,504 in our catalogue, hereafter referred to as ``The Seahorse Galaxy'' on account of its 1.5\,GHz radio morphology) is an extended source ($\mathrm{LAS}=1\farcs0$), the brightest component of which overlies the highly dust-obscured nuclear region of an $i=22.3^\mathrm{mag}$ merging S{\sc cd} galaxy at $z=1.224$ \citep{Barger2014:MaxSFR}. We measure total flux densities of $S_{\rm 1.5\,GHz}=174.0\pm5.6\,\mu$Jy and $S_{\rm 5.5\,GHz}=46.2\pm4.8\,\mu$Jy, respectively using our resolution-matched 1.5\,GHz and 5.5\,GHz maps, from which we find that the Seahorse has a low frequency spectral index which is consistent with aged synchrotron emission ($\alpha^{\rm 5.5\,GHz}_{\rm 1.5\,GHz}=-1.02\pm0.08$).

The Seahorse is the most likely radio counterpart to the SCUBA $850\,\mu$m source, HDF\,850.7 \citep{serjeant03}. We show \emerge\ radio images of this source in Fig.\,\ref{fig:ant_thumbs}. The total stellar mass of the merging system is estimated from SED-fitting to be $(9.5\pm 0.1)\times 10^{10}\,\mathrm{M_{\odot}}$ \citep{Skelton2014;3DHST}. The extended radio emission of the Seahorse overlies two bright optical components running to the south into a tidal tail. Combining our resolution-matched 1.5\,GHz and 5.5\,GHz maps, we create a spectral index map for the Seahorse, measuring a moderately steep ($\alpha\sim-0.7$) spectral index across the bright component, which steepens to $\alpha\sim-1.0$ as the extended radio component follows the red tail of the merging system. 

The Seahorse also lies within the GOODS-N Jansky VLA 10\,GHz Pilot Survey area \citep{murphy17}. Only the brightest component seen by \emerlin\ at 1.5\,GHz is detected at 10\,GHz, overlying the optically-obscured region and suggesting that the extended radio emission in the \emerlin-only image (whose $0\farcs 31\times 0\farcs 21$ PSF is similar to the $0\farcs 27\times0\farcs 23$ PSF of the 10\,GHz image) is the result of dust-obscured, spatially-extended star-formation rather than the blending of compact cores from the two progenitor galaxies in this merging system (Fig.\,\ref{fig:ant_thumbs}). \citet{murphy17} measure a flux density for The Seahorse of $S_{\rm 10\,GHz}=36.71\pm 0.06\,\mu$Jy. The 5.5-to-10\,GHz spectral index is therefore $\alpha^{\rm 10\,GHz}_{\rm 5.5\,GHz}=-0.38\pm0.25$, which is considerably flatter than the 1.5-to-5.5\,GHz spectral index measured previously, and is consistent with spectral flattening due to increasing thermal emission at higher frequencies.  

Our interpretation of the radio structure in the Seahorse is therefore that it is dominated by intense star-formation taking place within the very dusty regions of the merging system which produces obscuration and reddening in the optical bands. This radio emission in turn likely traces the regions from which the prodigious far-IR luminosity originates, owing to the FIRRC. The brightest component, which is detectable from 1.5--10\,GHz appears to have a flatter radio spectrum (due to the increased spatial density of H\textsc{ii} regions) than the surrounding material, which is undetected at 10\,GHz (and hence likely has a steeper spectrum tracing a synchrotron halo around the central starburst).

The Seahorse galaxy system illustrates the advantages of high angular resolution imaging at $\sim$GHz radio frequencies, where the older radio emitting plasma is more easily detected than at higher frequencies due to its spectral properties. Observing at $\nu_{\rm obs}\gtrsim 10$\,GHz with the VLA provides the required resolution to resolve such systems, but suffers from strong spectral selection effects which must be understood and disentangled before meaningful comparisons with samples selected in the GHz-window can be made.

\subsection{The redshift and luminosity distributions of \emerge\ DR1 sources}
To provide added value to the \emerge\ DR1 catalogue, we match our radio source list to the multi-band optical/infrared catalogue of \textit{HST} WFC3-selected sources compiled by the 3D-HST team \citep{Skelton2014;3DHST}. To check the astrometric accuracy of the 3D-HST catalogue we take the VLA source positions of the brightest 100 radio sources common to both the \emerge\ DR1 survey region and 3D-HST \textit{HST} WFC3 mosaic images, and stack in each of the F105W, F125W and F160W images at these source positions. We fit the stacked images with a two-dimensional Gaussian and measure the offsets in the fitted centroids from the centre of the thumbnail images (which are centred on the VLA source positions). We measure small ($\delta\theta\lesssim 0.2''$) linear offsets in RA. To correct for this, we apply a linear shift to the 3D-HST catalogue in RA, corresponding to the mean offset ($\delta\theta=0.1267''$).

With this shift applied we find optical counterparts to 587 of our 848 VLA-detected \emerge\ sources (69\%) within a $\sim 1\farcs 5$ error circle, providing redshift information and allowing both the luminosities and linear sizes of our radio-selected sample to be measured. Of the 261 \emerge\ DR1 sources without optical counterparts, 235 were found to lie outwith the footprint of the \textit{HST} F125W image which defines the survey area of 3D-HST \citep{Skelton2014;3DHST}. There are therefore 26 \emerge\ sources detected above $4.8\sigma$\ at 1.5\,GHz which lie within the 3D-HST survey area and which do not have counterparts in the 3D-HST source catalogue, an optical non-detection rate of $4.2\%$.

\begin{figure*}
    \centering
    \includegraphics[width=\textwidth]{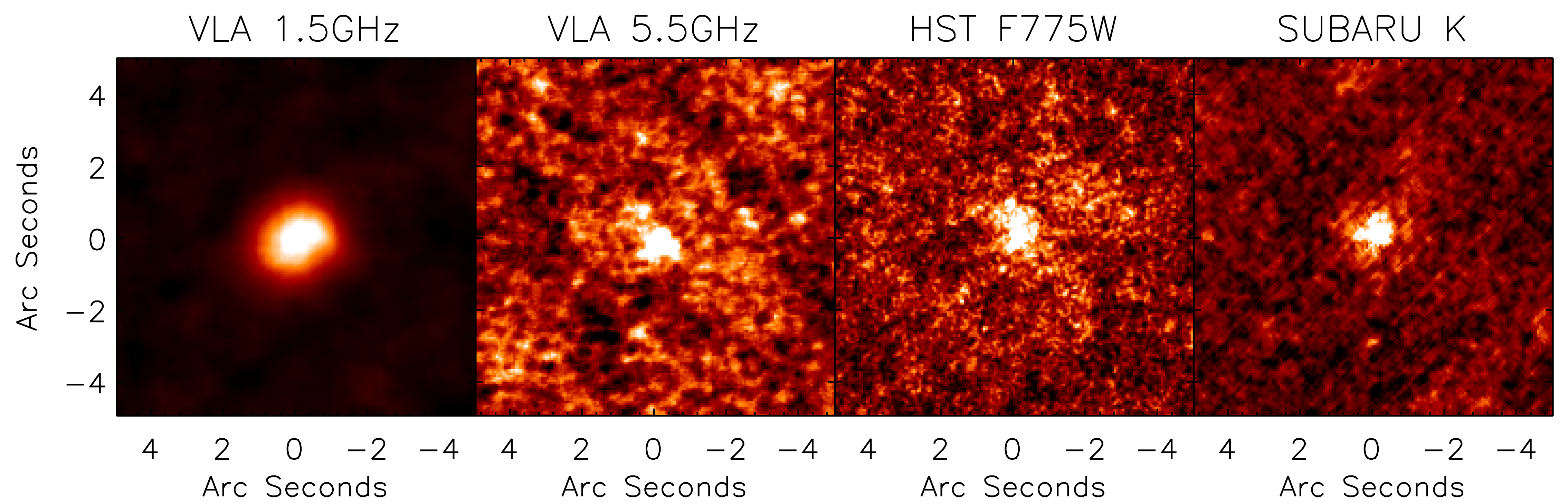}
    \caption{Stacked thumbnail images of 26 \emerge\ DR1 sources with VLA 1.5\,GHz detections above $4.8\times\sigma$, but no reported optical counterparts in the 3D-HST catalogue of \citet{Skelton2014;3DHST}. We fit the radio emission in the two stacked radio thumbnail images using Gaussian source components, measuring flux densities of $S_{\rm 1.5\,GHz}=29\pm 1\,\mu$Jy and $S_{\rm 5.5\,GHz}=11\pm 3\,\mu$Jy, and measure I and K-band magnitudes of $I=26.07\pm2.33$ and $K=24.01\pm0.62$ via \textit{HST} F775W and SUBARU K-band imaging from \citet{Skelton2014;3DHST} and \citet{kajisawa11}, respectively. The detection of emission in all four stacked thumbnails highlights that the 26 \emerge\ sources which lack counterparts in the 3D-HST optical catalogue are (on average) real sources, associated with red ($(I-K)=2.06\pm0.19$), faint ($K\sim24$) host galaxies.}
    \label{fig:stack}
\end{figure*}

To establish whether these 26 optically-blank radio sources are real or spurious, we extract thumbnail images at their measured radio positions in the VLA 1.5\,GHz and 5.5\,GHz radio images \citep[cf][]{beswick08} and \textit{HST} F775W (I-band) and Subaru K-band near-IR images, and stack the 26 thumbnail images in each waveband using a median stacking algorithm \citep[e.g.\ ][]{thomson17}. The stacked thumbnail images are shown in Fig.\,\ref{fig:stack}. By fitting Gaussian source components to the two stacked radio thumbnails using the {\sc casa} task {\sc imfit} we measure median radio flux densities of $S_{\rm 1.5\,GHz}=29\pm 1\,\mu$Jy and $S_{\rm 5.5\,GHz}=11\pm 3\,\mu$Jy. To measure median $AB$ magnitudes from the stacked optical thumbnails, we perform aperture photometry in Source Extractor \citep{bertin96} using a 1\farcs5 aperture and zero-point offsets of 25.671 \citep{Skelton2014;3DHST} and 26.0 \citep{kajisawa11} for the \textit{HST} F775W I-band and Subaru K-band images, respectively. We measure median magnitudes of $K=24.01\pm0.62$ and $I=26.07\pm2.33$. We detect significant emission in the four stacked thumbnail images, confirming that on average the 26 optically-undetected \emerge\ sources are real, albeit faint and red: $K\sim 24$ and $(I-K)=2.06\pm0.19$. This combination of radio flux densities and optical colours is consistent with emission from high-redshift ($z>2$) dust-obscured star-forming galaxies which are frequently missed in even the deepest optical studies \citep[e.g. ][]{smail02}.

\begin{figure}
    \centering
    \includegraphics[width=\columnwidth]{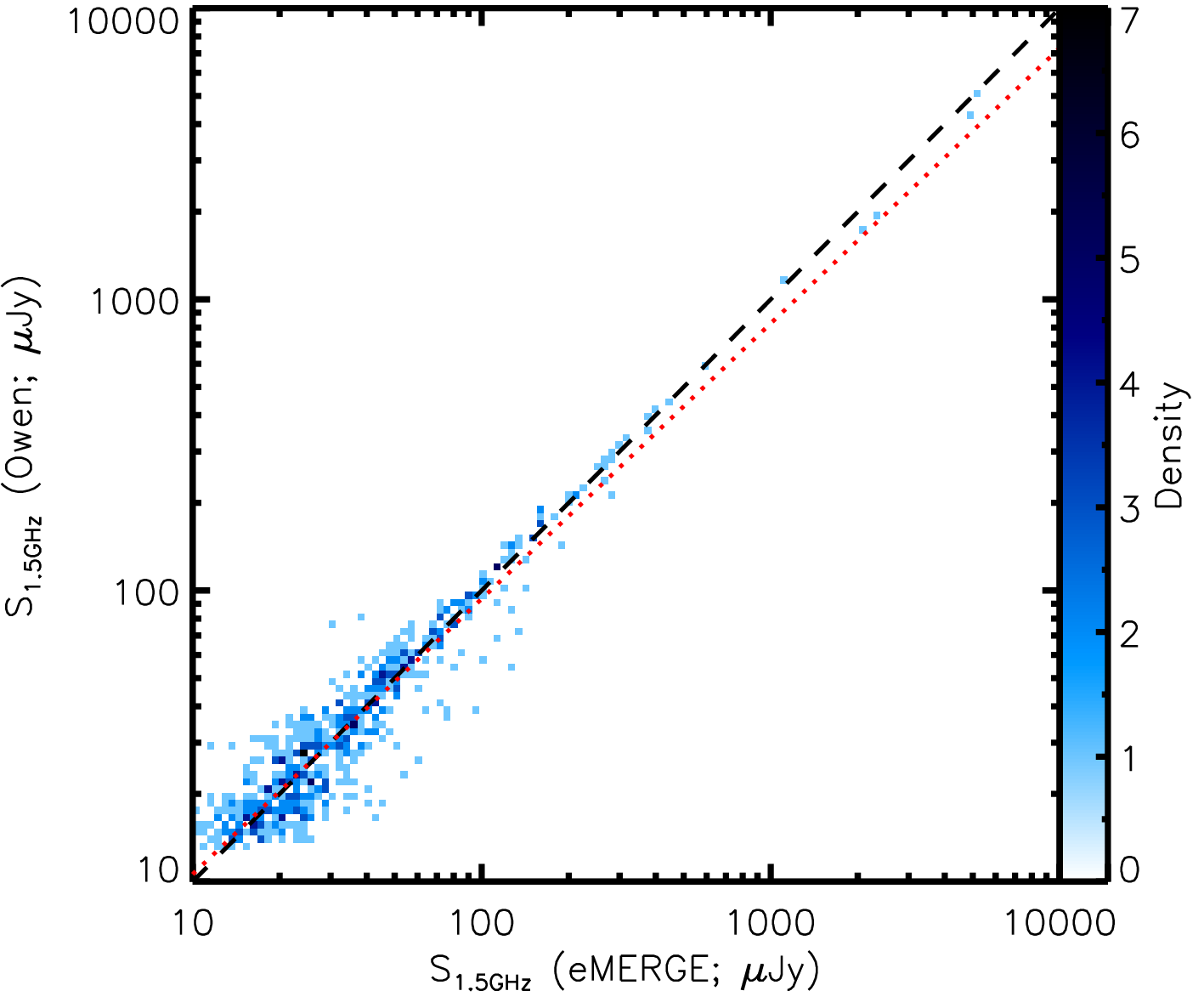}
    \caption{Flux density comparison between radio source components detected in the VLA 1.5\,GHz study of \citet{owen18} and those detected in our independent reduction and analysis of the same observations. We show our results as a density plot, with 2-dimensional bins of width $\log_{10}(S_{\rm 1.5\,GHz}/\mu{\rm Jy})=0.025$. A colour bar on the right hand side of the plot indicates the number of sources in each flux bin. We show the 1-to-1 relation as a dashed black line, along with the log-linear best fit, $S_{\rm 1.5\,GHz, Owen} = 0.95S_{\rm 1.5\,GHz, eMERGE}+0.08$, which is shown as a dotted red line. We have a tendency to measure slightly higher flux densities for our source components with respect to the flux densities presented in the catalogue of \citet{owen18}; we believe this result is due to the source-fitting methodology of {\sc pybdsf}, which uses wavelet decomposition to ``mop up'' extended emission which is not well fit by Gaussian source components.}
    \label{fig:muxlow_owen}
\end{figure}

To provide an independent check of our data reduction, imaging and cataloguing strategies, we compare the \emerge\ DR1 VLA source flux densities against those reported in the VLA GOODS-N catalogue of \citet{owen18}, whose analysis was based on an independent reduction of the same raw telescope data. Our imaging strategy differs from that used by \citet{owen18} in terms of data weights and imaging algorithms used \citep[e.g. ][uses the {\sc casa} {\sc tclean} package with multi-scale clean, $w$-projection and a Briggs {\sc robust} value of 0.5 whereas we use {\sc wsclean} with $w$-stacking and natural weighting]{owen18}. Moreover, the fields of view of the two VLA images differ slightly: \citet{owen18} images a circular field $18\arcmin$\ in diameter, and achieves a noise level near the centre of the field of $\sigma_{\rm 1.5\,GHz}=2.2\mu$Jy\,beam$^{-1}$ from 39\,hours of observations, detecting 795 radio sources down to $4.5\times\sigma_{\rm rms}$. As previously discussed (\S\,\ref{sect:data-vla}), the \emerge\ DR1 survey area is a $15\arcmin\times 15\arcmin$ square, and by including both the \citet[][post-upgrade]{owen18} and \citet[][pre-upgrade]{richards00} VLA observations in our imaging run, we achieve a noise level of $\sigma_{\rm 1.5\,GHz}=2.04\,\mu$Jy\,beam$^{-1}$ at the centre of the field. Of the 795 sources in the \citet{owen18} catalogue, 664 lie within the \emerge\ DR1 survey area. In turn, 812/848 \emerge\ DR1 sources lie within the footprint of the \citet{owen18} catalogue. We cross-match the \citet{owen18} and \emerge\ DR1 source catalogues using a 1\farcs5 matching radius (corresponding to the VLA 1.5\,GHz synthesized beam), finding 602 sources in common. This implies that within the area common to both studies there are 17 \emerge\ DR1 sources which are not in the \citet{owen18} catalogue, and 62 radio sources in the \citet{owen18} catalogue which are not in the \emerge\ DR1 catalogue. However, this 62 includes 24 extended ($>2$--$3$\arcsec) sources which visual inspection confirmed \textit{are} in fact in the \emerge\ catalogue, albeit with recorded source positions (determined by {\sc pybdsf}) which are $>1\farcs5$ away from the position determined by the \aips\ {\sc sad} routine used by \citet{owen18}. The remaining 38 sources are relatively low significance ($\langle{\rm S/N}\rangle=5.2$) detections in the \citet{owen18} catalogue, and the differing imaging and source identification strategies used may be enough to explain their non-detections in \emerge. We show a comparison between the \citet{owen18} and \emerge\ DR1 VLA integrated flux densities of 602 sources in Fig.\,\ref{fig:muxlow_owen}. We perform a linear least-squares fit to these flux densities, measuring $\log_{10}(S_{\rm 1.5\,GHz; Owen}/\mu{\rm Jy}) = 0.946\times\log_{10}(S_{\rm 1.5\,GHz; eMERGE}/\mu{\rm Jy}) + 0.079$. We believe this modest excess of flux in the \emerge\ catalogue with respect to the catalogue of \citet{owen18} can be partly explained by the different source-fitting methodologies: the {\sc sad} routine in \aips\ used by \citet{owen18} fits Gaussian models to detected source components using a least-squares method, whereas (as previously discussed) the {\sc atrous\_do} module in {\sc pybdsf} supplements this source-fititng with a wavelet decomposition module to capture the residual emission around extended sources which is not accounted for by Gaussian source fitting alone. The sum of the integrated flux densities of these 602 sources in the \emerge\ DR1 catalogue is 45.83\,mJy, 3.8\% higher than the sum of the integrated flux densities of the same sources in the \citet{owen18} catalogue (44.15\,mJy). The median flux densities of these 602 sources, however, are $30.7\pm1.6\,\mu$Jy and $30.0\pm1.1\,\mu$Jy in \emerge\ DR1 and the catalogue of \citet{owen18}, respectively, which are consistent within the measurement errors. We therefore conclude that there are no significant offsets in our overall flux scale with respect to \citet{owen18}, and that our source catalogues are consistent to within the overall flux scale calibration uncertainties of the VLA\footnote{\url{https://science.nrao.edu/facilities/vla/docs/manuals/oss/performance/fdscale}}.

\begin{figure*}
\begin{centering}
\includegraphics[width=0.95\textwidth]{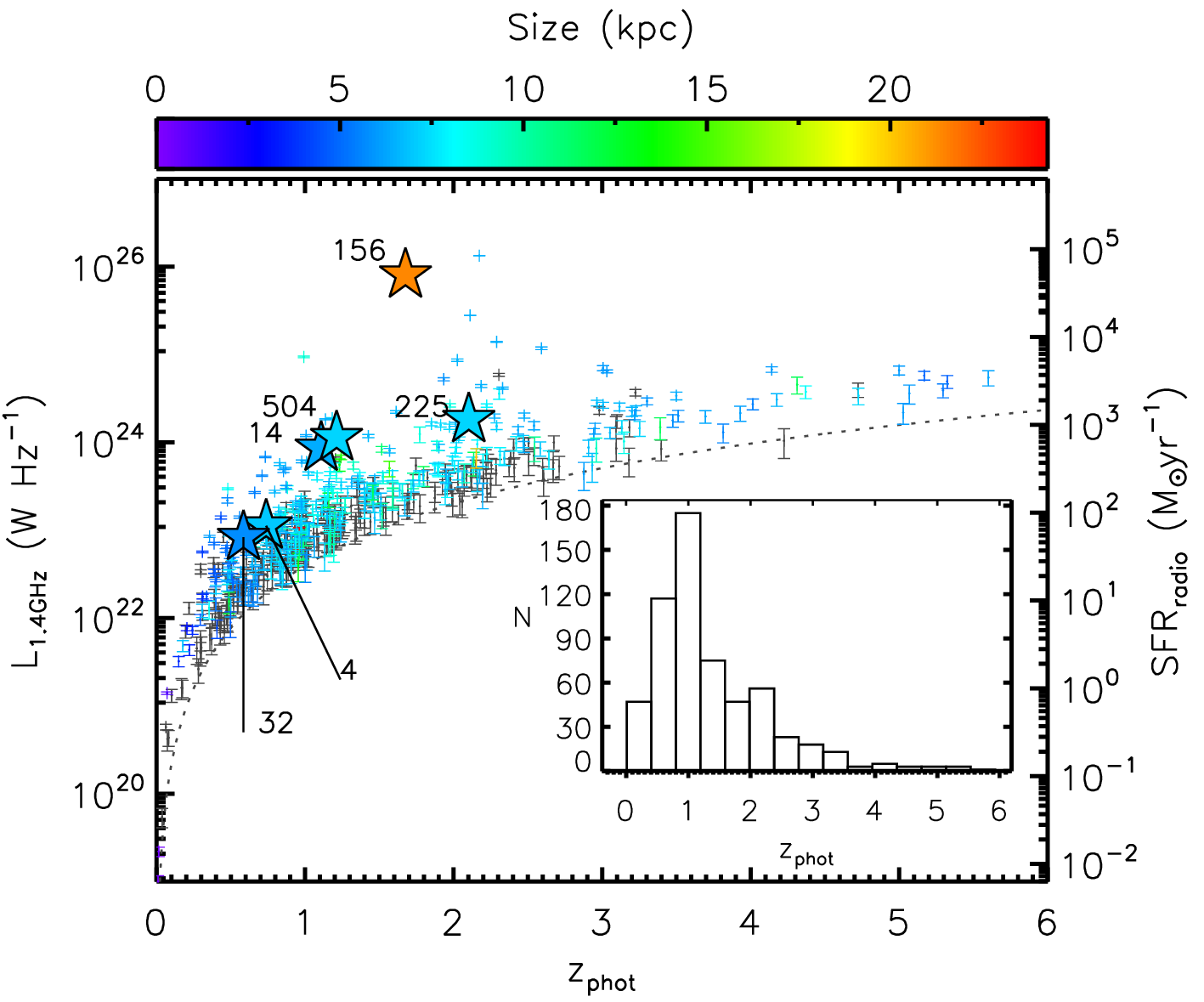}
\vspace*{-4mm}
\caption{\textit{Main panel: }The luminosity--redshift plane for \emerge\ DR1, including the 587 radio-detected sources with optical counterparts within $1\farcs5$ from the 3D-HST catalogue \citep{Skelton2014;3DHST}. We measure rest-frame $L_{\rm 1.4\,GHz}$ from our observed-frame 1.5\,GHz flux densities using this redshift information, along with: (i) the measured radio spectral index ($\alpha^{\rm 5.5\,GHz}_{\rm 1.5\,GHz}$) for sources detected in both \emerge\ bands; (ii) $\alpha=-0.8$ for sources which are non-detected at 5.5\,GHz (provided that spectral index is consistent with the 5.5\,GHz non-detection); (iii) $\alpha < -0.8$, if required by the corresponding $3\times\sigma_{\rm 5.5\,GHz}$ upper-limits. To illustrate our sensitivity to SFR as a function of redshift, we use the 1.4\,GHz-to-SFR conversion factor of \citet{murphy11}, which highlights our ability to detect high-SFR systems at high-redshift (i.e.\ ${\rm SFR}\sim 100$--$1000$\,M$_\odot$\,yr$^{-1}$ at $z\sim 2.5$). Points are colour-coded by the fitted radio sizes (if measured; see \S\,\ref{sect:sizes}), with sources which lack a significant size measurement coloured in charcoal. We highlight 6 of the illustrative sources shown in Fig.\,\ref{fig:eMERGE_5maps} with large star symbols. \textit{Inset: } The photometric redshift distribution of \emerge\ DR1 peaks at $\langle z \rangle = 1.08\pm 0.04$, with a tail (accounting for $\sim 15\%$ of the sample) lying between $z=2.0$--$5.6$ \citep{Skelton2014;3DHST}.}
\label{fig:luminosity_redshift_colour}
\end{centering}
\end{figure*}

\begin{figure}
    \centering
    \includegraphics[width=\columnwidth]{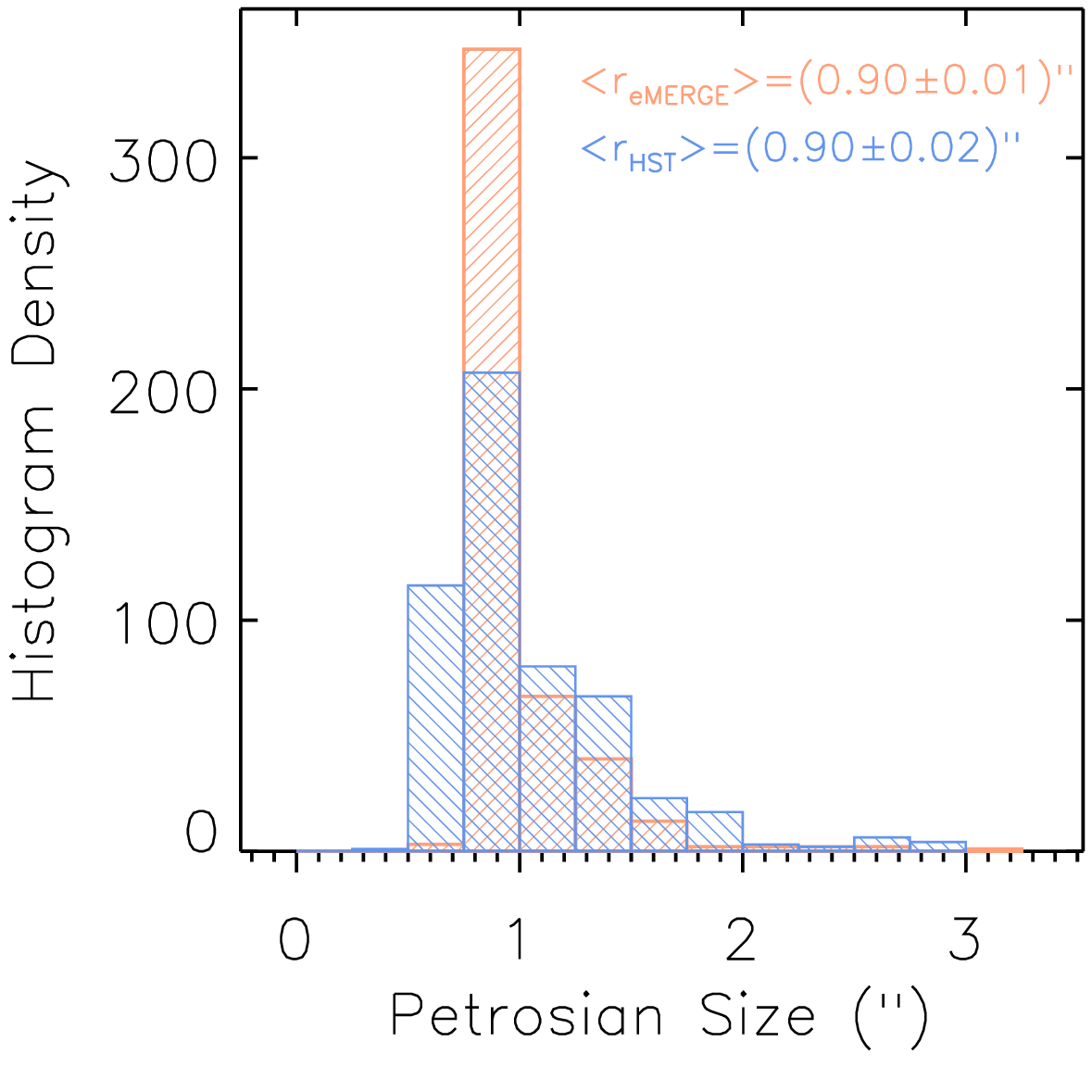}
    \caption{Histogram of optical and radio angular sizes of \emerge\ sources. From our parent catalogue of 848 VLA 1.5\,GHz-detected sources, we measure deconvolved radio petrosian sizes for 479 galaxies (56\% of the sample) and deconvolved optical sizes (from a stacked three-band \textit{HST} CANDELS F606W, F814W and F850LP image, smoothed to match the 0\farcs8 beam of our \emerge\ ``maximum sensitivity'' radio image) for 525 galaxies (62\% of the sample). Galaxies without size measurements include optically-blank radio sources (3\%), and sources whose size measurements are consistent with being up-scattered point sources (either because the light profiles of these sources are noise-like, or because they lie within a few arcseconds of a much brighter source, and the Petrosian size cannot be disentangled from the blended light profile: 35\%). The median 1.5\,GHz radio and optical sizes in \emerge\ are $\langle r_{\rm eMERGE}\rangle = 0\farcs90\pm0\farcs01$\ and and $\langle r_{HST}\rangle = 0\farcs90\pm0\farcs02$, respectively. Note that these histograms include sources both with and without photometric redshift information from the 3D-HST catalogue.}
\label{fig:size_histogram}
\end{figure}

From the SED fitting work of \citet{Skelton2014;3DHST}, the \emerge\ DR1 sample has a median photometric redshift of $z_{\rm phot}=1.08\pm0.04$, with a tail of sources ($\sim 10$\% of the sample) lying at $z=2.5$--$6$ (see Fig.\,\ref{fig:luminosity_redshift_colour}). We use these photometric redshifts to \textit{k}-correct our observed-frame 1.5\,GHz flux densities to rest-frame 1.4\,GHz, and measure radio luminosity densities $L_{\rm 1.4\,GHz}$ via the Equation 1 of \citet{thomson19b}, though with an additional correction $A\equiv (1.40/1.51)^{-\alpha}$ which accounts for the slight offset in frequency between our observed-frame 1.51\,GHz observations and the observed-frame 1.4\,GHz, which is the central frequency most commonly associated with L-band radio observations:

\begin{center}
  \begin{equation}
    L_{\rm 1.4\,GHz, rest} = 4\pi D_{L}^{2}AS_{\rm 1.51\,GHz,rest}(1+z)^{-1-\alpha}
  \end{equation}
\end{center}

For sources detected at both 1.5\,GHz and 5.5\,GHz, we \textit{k}-correct using the measured radio spectral index, $\alpha^{\rm 5.5\,GHz}_{\rm 1.5\,GHz}$, and for sources detected at 1.5\,GHz but not at 5.5\,GHz we use either $\alpha=-0.8$ (if consistent with the 5.5\,GHz non-detection), or a steeper spectral index if required by the $3\times\sigma_{\rm 5.5\,GHz}$ upper-limit.

The luminosity/redshift distribution of our sources is shown in Fig.\,\ref{fig:luminosity_redshift_colour}. To illustrate our sensitivity to star-formation as a function of redshift, we convert these radio luminosities in to equivalent star formation rates using the relation found in \citet{murphy11}, i.e.\

\begin{center}
\begin{equation}
\log_{10}({\rm SFR}/{\rm M}_\odot {\rm yr}^{-1}) = \log_{10}(L_{\rm 1.4\,GHz}/{\rm erg\,s}^{-1}\,{\rm Hz}^{-1}) - 28.20
\label{eq:murphy_sfr}
\end{equation}
\end{center}

\noindent which assumes a \citet{kroupa01} stellar initial mass function (IMF), integrated between stellar mass limits of $0.1$--$100$\,M$_{\odot}$.

While we emphasise that it is highly unlikely that any radio-selected galaxy sample at high-redshift will be entirely dominated by star-formation, we see in Fig.\,\ref{fig:luminosity_redshift_colour} that the \emerge\ DR1 maps are sufficiently sensitive to detect radio emission from a combination of AGN and high-SFR systems, such as SMGs (${\rm SFR}\sim 100$--$1000$\,M$_\odot$\,yr$^{-1}$), at least out to $z\sim 2.5$. For $z\gtrsim 3$, the strong positive \textit{k}-correction in the radio bands biases our flux-limited 1.5\,GHz sample toward radio sources which are an order of magnitude more luminous than typical SMGs; these high-power, high-redshift radio systems are almost certainly an AGN-dominated population. We defer detailed classification of the \emerge\ radio source population to future publications.

\subsection{The radio sizes of SFGs and AGN from $z=1$--$3$}\label{sect:sizes}

\begin{figure*}
\begin{centering}
\includegraphics[width=\textwidth]{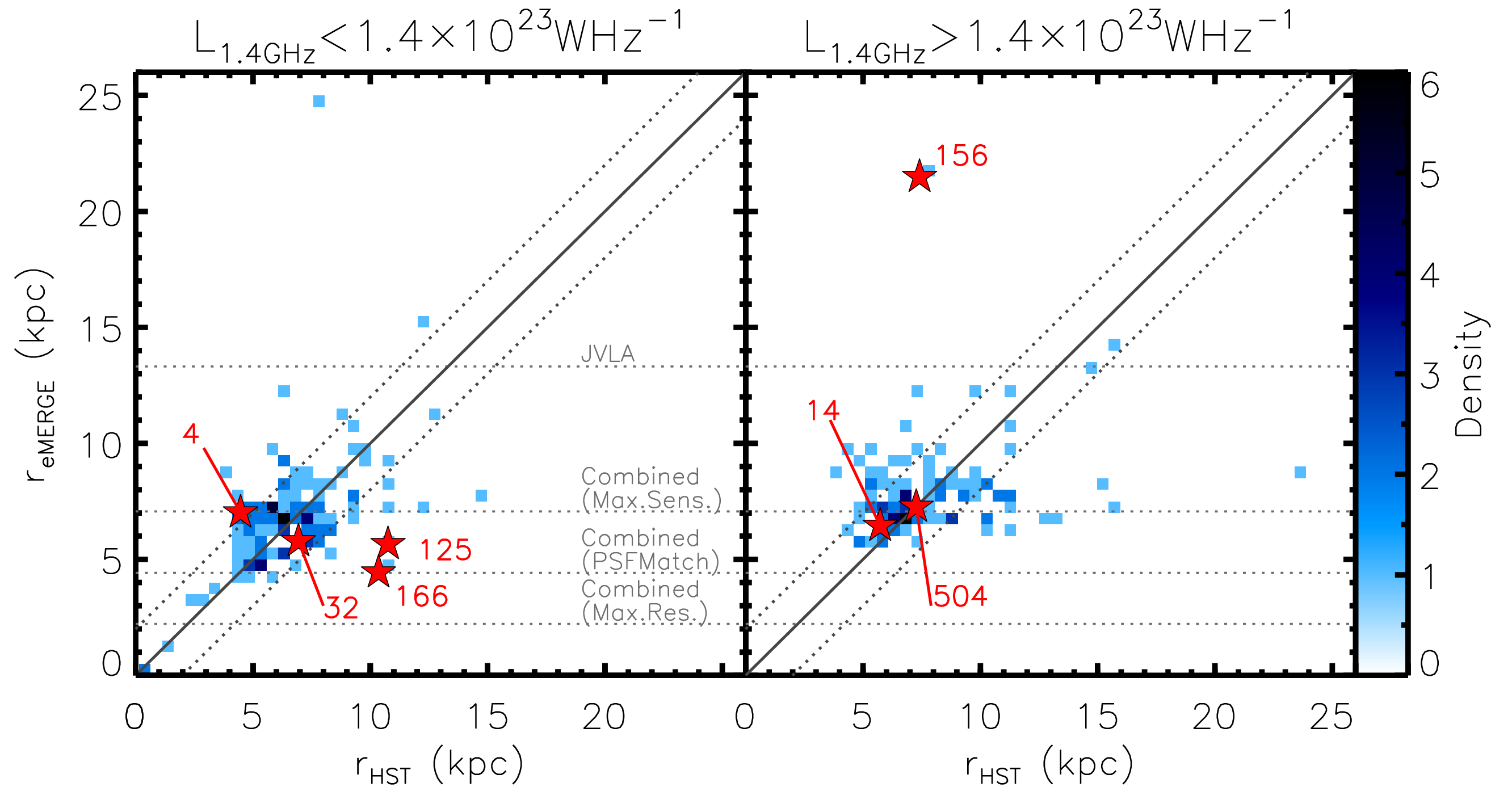}
\vspace*{-4mm}
\caption{The deconvolved radio and rest-frame optical size distribution of 248 \emerge\ radio sources with Petrosian size measurements and photometric redshift information from the 3D-HST survey \citep{Skelton2014;3DHST}. Linear radio sizes are measured from our ``Maximum Sensitivity'' combined-array image, which has an angular resolution of $\sim 0\farcs8$, and optical sizes are measured from a stacked three-band (F606W, F814W, F850LP) \textit{HST} CANDELS image, after smoothing to the same resolution as our radio map. The distribution of linear sizes is shown as two density plots, in bins of width $0.5\,{\rm kpc}\times 0.5\,{\rm kpc}$, with horizontal lines showing the effective linear size of the PSFs of the suite of \emerge\ images at the median redshift of the sample ($<z>=1.08\pm0.04$). \emerge\ allows us for the first time to measure the angular sizes of ``normal'' galaxies at $z\sim1$ at 1.5\,GHz, revealing a mean radio-to-optical size ratio of $1.02\pm 0.03$. Large star symbols represent the individual sources in Fig.\,\ref{fig:eMERGE_5maps} for which both radio and optical size measurements are available. \textit{Left: }\emerge\ radio versus \textit{HST} optical sizes for 124 galaxies below the median radio luminosity of the sample ($L_{\rm 1.4\,GHz}<1.4\times 10^{23}$\,W\,Hz$^{-1}$). The density plot peaks around a median $1.04\pm0.05$, based on a radio size of $r_{\rm eMERGE}=6.74\pm0.23$\,kpc and optical size of $r_{\rm HST}=6.45\pm0.20$\,kpc. \textit{Right: } Radio versus optical size density plot for the brighter half of the sample ($L_{\rm 1.4\,GHz}>1.4\times 10^{23}$\,W\,Hz$^{-1}$). The median \emerge-to-optical size ratio for brighter radio sources is $1.00\pm0.04$, based on a radio size $r_{\rm eMERGE}=7.73\pm0.19$\,kpc and optical size of $r_{\rm HST}=7.73\pm0.27$. In both subsamples, therefore, the radio emission appears to trace similar scales to the optical stellar light, suggesting a source population which is not dominated by jetted radio AGN. At higher radio powers there is weak evidence ($\sim 1\sigma$) of a lower radio-to-optical size ratio than at weaker radio powers, which may be indicative of a radio source population in which compact nuclear AGN emission begins to play a more prevalent role.}
\label{fig:sizeradio_sizehst}
\end{centering}
\end{figure*}

To measure the (sub-arcsecond) resolved radio properties of \emerge\ DR1 sources, we use the VLA source catalogue to provide positional priors and then measure the Petrosian radii \citep[$R_P$; following][]{petrosian1976surface, wrigley16} of these sources in the higher-resolution combined-array \emerlin+VLA and \emerlin-only images. We define $R_{\rm P}$ as the radius $r$ at which the local surface brightness profile, $I(R)$, equals $0.4\times$ the mean surface brightness within $R_{\rm P}$, $\langle I\rangle_R$ \citep[e.g.\ ][]{graham05}.

As discussed in \S\,\ref{sect:imaging}, the suite of five \emerge\ 1.5\,GHz DR1 images offers a sliding-scale in both angular resolution and surface brightness sensitivity between the extremes of VLA-only and \emerlin-only imaging. To provide a set of representative source size measurements for our high-redshift galaxy sample, we focus our analysis on the \emerlin+VLA ``Maximum Sensitivity'' image (Table\,\ref{tab:images}). This image has an angular resolution of $0\farcs89\times 0\farcs78$ and an rms sensitivity of $\sigma_{\rm rms}=1.71\,\mu$Jy\,beam$^{-1}$, (corresponding to a linear scale of $\sim 7$\,kpc and a $3\sigma$ point source star formation rate sensitivity of $15$\,M$_\odot$\,yr$^{-1}$\,beam$^{-1}$ at $z=1$, using Equation\,\ref{eq:murphy_sfr}). This image is thus well suited to providing canonical size measurements for star-forming galaxies at high-redshift, whose typical optical angular sizes are $\sim 5$--$10$\,kpc \citep{williams10, vanderwel14, rujopakarn16}.

We measure the uncertainties on the individual Petrosian size estimates using a Monte Carlo process, wherein for each source, for each annulus we perturb every pixel by a value drawn from a Gaussian distribution of fluxes whose width is equal to the local rms, and re-fit the profile. We repeat this process 100 times per annulus, and define the $\pm1\sigma$\ error on the fitted size which results from this process for each source from the range of minimum/maximum Petrosian sizes allowed by this process. These size errors -- along with profiles for each source -- will be presented along with the source catalogue in a forthcoming publication (Thomson \etal, in prep). 

In order to provide a consistent comparison between the radio and optical size distributions, we smooth the \textit{HST} CANDELS F606W, F814W and F850LP images to match the resolution of the  maximum-sensitivity \emerge\ image, and then co-add these images in order to provide a high signal-to-noise, broad-band optical image. We then fit Petrosian optical sizes using the same methodology as was employed in our radio imaging. We show the size histograms from fitting to our radio and stacked optical images in Fig.\,\ref{fig:size_histogram}. Of the 587 \emerge\ DR1 sources with optical counterparts in the 3D-HST catalogue, 312 both have the photometric redshift information needed to derive linear source sizes, and yield convergent Petrosian size measurements in both the radio and the optical bands. The remaining 275 sources either lack a photometric redshift measurement, have too little S/N to fit a resolved profile in one or both images, or lie within crowded fields, and hence $I(R)>0.4\times\langle I\rangle_R$ for all physically plausible sizes (i.e.\ $\lesssim$50\,kpc).

To test whether these 312 deconvolved size measurements represent real source structure, or whether they represent spurious fitting of point sources, we create a simulated image (with Gaussian noise) and inject 20,000 point sources with signal-to-noise ratios between $0<{\rm S/N}<20$, and then convolve this with the combination image dirty beam. We then perform Petrosian size fitting on this simulated image at the positions of the known point sources. Unsurprisingly, we find that lower signal-to-noise point sources have a greater tendency to be up-scattered in size than higher signal-to-noise point sources. Following \citet{bondi08} and \citet{thomson19b}, we parameterise this size ``up-scattering'' as a function of signal-to-noise by measuring the envelope in size versus signal-to-noise below which 99\% of the simulated point sources lie. We determine that $\lesssim 1\%$ of point sources scatter above an envelope of $\log(R_{\rm P}/{\rm arcsec}) = -1.05\log({\rm S/N})-0.25$. Applying this envelope to the Petrosian size measurements derived from our real data, we find that 64/312 source sizes are consistent with being unresolved at $0\farcs7$ resolution. The remaining 248 sources represent the largest high-redshift galaxy sample to date with resolved kpc-scale size measurements at 1.5\,GHz: this sample is poised to expand with our forthcoming, deeper, wider-area DR2 data release. 

We show a comparison between the radio and optical Petrosian sizes of these 248 \emerge\ galaxies in Fig.\,\ref{fig:sizeradio_sizehst}. The mean radio-to-optical size ratio of sources detected in both images is $1.02\pm0.03$ -- where the uncertainty quoted is the standard error (i.e. $\sigma/\sqrt{n}$) -- implying that the radio emission traces the rest-frame UV stellar light. Splitting this sample near the median radio luminosity ($L_{\rm 1.4\,GHz}\sim 1.4\times 10^{23}$\,W\,Hz$^{-1}$) we measure a radio-to-optical size ratio of $1.04\pm 0.05$ for the fainter half of the sample and $1.00\pm0.04$ for the brighter half. The median radio sizes are $7.7\pm0.2$\,kpc and $6.7\pm0.1$\,kpc above and below the median luminosity, respectively. This increase in radio size with radio luminosity is consistent with the findings of \citet{bondi18}, whose study on the size evolution of the 3\,GHz-selected VLA-COSMOS sample \citep{smolcic17} uncovered similar behaviour for both the radio-loud AGN and radio-quiet AGN in their sample. Our near-unity radio-to-optical size ratio is in tension with a results from the VLA COSMOS 3\,GHz study \citep{jimenez-andrade19}, whose median radio size is $\sim 1.3$--$2\times$\ smaller than the optical/UV continuum emission which traces the stellar component. Given that the VLA COSMOS 3\,GHz and \emerge\ 1.5\,GHz maximum sensitivity images are of comparable angular resolution ($\theta_{\rm COSMOS}=0\farcs75$, cf $\theta_{\rm eMERGE}=0\farcs84$), it is unlikely this discrepancy in radio-to-optical size ratios is a result of resolution effects, but rather reflects real differences in the physical scales of processes emitting at different radio frequencies. As previously suggested by \citet{murphy17} and \citet{thomson19b}, these differences may include frequency-dependent cosmic ray diffusion and/or may be due to the increasing thermal fraction at higher rest-frame radio frequencies, revealing time lags between the production of free-free and synchrotron emission in star-forming galaxies \citep[e.g.\ ][]{bressan02, thomson14, gomez-guijarro19}.

20 sources ($\sim 8$\% of the 248 \emerge\ galaxies with fitted optical and radio size information) have radio-to-optical size ratios which exceed 1.2. These include ID\,156, a wide-angle tailed radio source associated with a compact red elliptical host galaxy. There are 21 galaxies ($\sim 8$\% of the sample with size information) which have radio-to-optical size ratios smaller than 0.8. These include ID\,166, a face-on spiral galaxy with radio emission tracing star-formation down one of the spiral arms only (see Fig.\,\ref{fig:eMERGE_5maps} for details).

Recently, \citet{lindroos18} used the $uv$-stacking technique combined with S\'{e}rsic model fitting (with a fixed $n=1$) to measure the optical and radio size evolution of optically-selected star-forming galaxies from $z=0$--$3$ across a stellar mass range $M_\star \sim 10^{10.5}-10^{11}$\,M$_\odot$ using the (pre-2010) MERLIN and VLA GOODS-N data which are included as part of \emerge\ DR1. \citet{lindroos18} found that the median radio sizes become larger at lower redshift, and that they are on average $\sim 2\times$ smaller both than the optical sizes of the same stacked samples, and than the H$\alpha$ sizes of typical star-forming galaxies. They concluded that radio continuum emission therefore preferentially traces morphologically compact star formation, concentrated towards the centres of galaxies. We do not see this trend among the 248 \emerge\ sources for which we can measure accurate sizes in our maximum-sensitivity combination image (see Fig.\,\ref{fig:luminosity_redshift_colour}). However we note that the PSF of this maximum-sensitivity combination image -- at half the size of the VLA-only PSF -- still only marginally-resolves structures which are $\sim6$\,kpc in size at $z\gtrsim 1$, and that around 49\% of our optically-detected sample do not have reliable size measurements in this map. If the radio emission in high-redshift galaxies is significantly more compact than even the $\sim 0\farcs8$ beam of our \emerge\ maximum sensitivity image, then we may see evidence of size-evolution in the higher-resolution (but lower surface brightness) \emerge\ DR1 images in future publications.

\section{Conclusions}\label{sect:conclusions}

In this paper we have described the motivation, design, data reduction and imaging strategies underpinning \emerge, a large legacy project combining \emerlin\ and VLA data at 1.5\,GHz and 5.5\,GHz (along with previously-obtained but newly re-processed observations from the pre-2010 MERLIN and VLA instruments). \emerge\ combines the long baseline capabilities of \emerlin\ with the high surface brightness sensitivity of the VLA to form a unique deep-field radio survey capable of imaging and studying the $\mu$Jy radio source population (i.e.\ star-forming galaxies and AGN at $z\gtrsim 1$) at sub-arcsecond angular resolution with high surface brightness sensitivity ($\sigma_{\rm 1.5\,GHz}\sim 1.5\,\mu$Jy\,beam$^{-1}$).

We have presented a description of the procedure for modelling the complicated \emerlin\ primary beam,  described post-processing steps which we applied to our data to correct for the deliterious effects of strongly variable unresolved sources within our target field, and described the imaging strategies necessary to seamlessly combine \emerlin\ and VLA data in the $uv$ plane in order to better the capabilities of either telescope individually.

We have shown some early science results from \emerge, including an analysis of the redshifts, radio luminosities and/or linear sizes of $\sim 500$ cosmologically-distant radio-selected sources. Our redshift distribution peaks at $\langle z \rangle = 1.08\pm 0.04$, with a tail ($\sim 15\%$ of the sample) lying at redshifts $z=2$--$5.6$. The sensitivity of \emerge\ DR1 is such that both AGN and starburst galaxies (${\rm SFR}=10^2$--$10^3$\,M$_\odot$\,yr$^{-1}$) are expected to be found in large numbers out to at least $z\sim 3$. We have highlighted the ability of \emerge\ to spatially-resolve high-redshift star-forming galaxies via an analysis of a $z=1.2$ dust-obscured SMG detected in three radio frequency bands (1.5\,GHz, 5.5\,GHz and 10\,GHz). We see evidence for significant size evolution in this source across the three frequency bands, with the 1.5\,GHz emission tracing scales roughly twice as large as those traced at 10\,GHz at comparable resolution.

This is intended as the first in a series of publications which will explore the full scientific potential of our suite of sensitive, high-resolution 1.5\,GHz and 5.5\,GHz images of the GOODS-N field as probes of star-formation and AGN activity in high-redshift source populations.

\section*{Acknowledgements}

APT, TWBM, RJB and MAG acknowledge support from STFC (ST/P000649/1). APT further acknowledges and thanks Robert Dickson, Peter Draper, David Hempston, Anthony Holloway and Alan Lotts for their extensive support managing the high-performance computing infrastructure which has been essential for the successful delivery of \emerge\ DR1. JFR acknowledges the financial assistance of the South African Radio Astronomy Observatory (SARAO) towards this research (\url{http://sarao.ac.za}). IP acknowledges support from INAF under the PRIN SKA/CTA project ``FORECaST'' and the PRIN MAIN STREAM project ``SAuROS''. E.I.\ acknowledges partial support from FONDECYT through grant N$^\circ$\,1171710. We thank the anonymous reviewer for their detailed and constructive referee report, which has undoubtedly improved the quality of the final manuscript. \emerlin\ is a National Facility operated by the University of Manchester at Jodrell Bank Observatory on behalf of STFC. The National Radio Astronomy Observatory is a facility of the National Science Foundation operated under cooperative agreement by Associated Universities, Inc. The authors acknowledge the use of the IRIDIS High Performance Computing Facility, and associated support services at the University of Southampton, in the completion of this work. We thank IRIS (\url{www.iris.ac.uk}) for provision of high-performance computing facilities. STFC IRIS is investing in the UK's Radio and mm/sub-mm Interferometry Services in order to improve data quality and throughput. The European VLBI Network is a joint facility of independent European, African, Asian, and North American radio astronomy institutes. Scientific results from data presented in this publication are derived from the following EVN project code(s): EG078B. The research leading to these results has received funding from the European Commission Horizon 2020 Research and Innovation Programme under grant agreement No. 730562 (RadioNet).




\bibliographystyle{mnras}
\bibliography{eMERGE_I}

\begin{thebibliography}{}
\makeatletter
\relax
\def\mn@urlcharsother{\let\do\@makeother \do\$\do\&\do\#\do\^\do\_\do\%\do\~}
\def\mn@doi{\begingroup\mn@urlcharsother \@ifnextchar [ {\mn@doi@}
  {\mn@doi@[]}}
\def\mn@doi@[#1]#2{\def\@tempa{#1}\ifx\@tempa\@empty \href
  {http://dx.doi.org/#2} {doi:#2}\else \href {http://dx.doi.org/#2} {#1}\fi
  \endgroup}
\def\mn@eprint#1#2{\mn@eprint@#1:#2::\@nil}
\def\mn@eprint@arXiv#1{\href {http://arxiv.org/abs/#1} {{\tt arXiv:#1}}}
\def\mn@eprint@dblp#1{\href {http://dblp.uni-trier.de/rec/bibtex/#1.xml}
  {dblp:#1}}
\def\mn@eprint@#1:#2:#3:#4\@nil{\def\@tempa {#1}\def\@tempb {#2}\def\@tempc
  {#3}\ifx \@tempc \@empty \let \@tempc \@tempb \let \@tempb \@tempa \fi \ifx
  \@tempb \@empty \def\@tempb {arXiv}\fi \@ifundefined
  {mn@eprint@\@tempb}{\@tempb:\@tempc}{\expandafter \expandafter \csname
  mn@eprint@\@tempb\endcsname \expandafter{\@tempc}}}

\bibitem[\protect\citeauthoryear{{Ashby} et~al.,}{{Ashby}
  et~al.}{2013}]{ashby13}
{Ashby} M.~L.~N.,  et~al., 2013, \mn@doi [\apj] {10.1088/0004-637X/769/1/80},
  \href {https://ui.adsabs.harvard.edu/abs/2013ApJ...769...80A} {769, 80}

\bibitem[\protect\citeauthoryear{{Baldi} et~al.,}{{Baldi}
  et~al.}{2018}]{baldi18}
{Baldi} R.~D.,  et~al., 2018, \mn@doi [\mnras] {10.1093/mnras/sty342}, \href
  {https://ui.adsabs.harvard.edu/abs/2018MNRAS.476.3478B} {476, 3478}

\bibitem[\protect\citeauthoryear{{Barger}, {Cowie}, {Sanders}, {Fulton},
  {Taniguchi}, {Sato}, {Kawara}  \& {Okuda}}{{Barger} et~al.}{1998}]{barger98}
{Barger} A.~J.,  {Cowie} L.~L.,  {Sanders} D.~B.,  {Fulton} E.,  {Taniguchi}
  Y.,  {Sato} Y.,  {Kawara} K.,   {Okuda} H.,  1998, \mn@doi [\nat]
  {10.1038/28338}, \href
  {https://ui.adsabs.harvard.edu/abs/1998Natur.394..248B} {394, 248}

\bibitem[\protect\citeauthoryear{{Barger} et~al.,}{{Barger}
  et~al.}{2014}]{Barger2014:MaxSFR}
{Barger} A.~J.,  et~al., 2014, \mn@doi [\apj] {10.1088/0004-637X/784/1/9},
  \href {https://ui.adsabs.harvard.edu/abs/2014ApJ...784....9B} {784, 9}

\bibitem[\protect\citeauthoryear{{Barger}, {Cowie}, {Owen}, {Hsu}  \&
  {Wang}}{{Barger} et~al.}{2017}]{barger17}
{Barger} A.~J.,  {Cowie} L.~L.,  {Owen} F.~N.,  {Hsu} L.~Y.,   {Wang} W.~H.,
  2017, \mn@doi [\apj] {10.3847/1538-4357/835/1/95}, \href
  {https://ui.adsabs.harvard.edu/abs/2017ApJ...835...95B} {835, 95}

\bibitem[\protect\citeauthoryear{{Bell}}{{Bell}}{2003}]{bell03}
{Bell} E.~F.,  2003, \mn@doi [\apj] {10.1086/367829}, \href
  {https://ui.adsabs.harvard.edu/abs/2003ApJ...586..794B} {586, 794}

\bibitem[\protect\citeauthoryear{{Bertin} \& {Arnouts}}{{Bertin} \&
  {Arnouts}}{1996}]{bertin96}
{Bertin} E.,  {Arnouts} S.,  1996, \mn@doi [\aaps] {10.1051/aas:1996164}, \href
  {https://ui.adsabs.harvard.edu/abs/1996A&AS..117..393B} {117, 393}

\bibitem[\protect\citeauthoryear{{Best}, {Kaiser}, {Heckman}  \&
  {Kauffmann}}{{Best} et~al.}{2006}]{best06}
{Best} P.~N.,  {Kaiser} C.~R.,  {Heckman} T.~M.,   {Kauffmann} G.,  2006,
  \mn@doi [\mnras] {10.1111/j.1745-3933.2006.00159.x}, \href
  {https://ui.adsabs.harvard.edu/abs/2006MNRAS.368L..67B} {368, L67}

\bibitem[\protect\citeauthoryear{{Beswick}, {Muxlow}, {Thrall}, {Richards}  \&
  {Garrington}}{{Beswick} et~al.}{2008}]{beswick08}
{Beswick} R.~J.,  {Muxlow} T.~W.~B.,  {Thrall} H.,  {Richards} A.~M.~S.,
  {Garrington} S.~T.,  2008, \mn@doi [\mnras]
  {10.1111/j.1365-2966.2008.12931.x}, \href
  {https://ui.adsabs.harvard.edu/abs/2008MNRAS.385.1143B} {385, 1143}

\bibitem[\protect\citeauthoryear{{Bhatnagar}, {Rau}  \& {Golap}}{{Bhatnagar}
  et~al.}{2013}]{bhatnagar2013}
{Bhatnagar} S.,  {Rau} U.,   {Golap} K.,  2013, \mn@doi [\apj]
  {10.1088/0004-637X/770/2/91}, \href
  {https://ui.adsabs.harvard.edu/abs/2013ApJ...770...91B} {770, 91}

\bibitem[\protect\citeauthoryear{{Biggs} \& {Ivison}}{{Biggs} \&
  {Ivison}}{2008}]{biggs08}
{Biggs} A.~D.,  {Ivison} R.~J.,  2008, \mn@doi [\mnras]
  {10.1111/j.1365-2966.2008.12869.x}, \href
  {https://ui.adsabs.harvard.edu/abs/2008MNRAS.385..893B} {385, 893}

\bibitem[\protect\citeauthoryear{{Bondi}, {Ciliegi}, {Schinnerer}, {Smol{\v
  c}i{\'c}}, {Jahnke}, {Carilli}  \& {Zamorani}}{{Bondi}
  et~al.}{2008}]{bondi08}
{Bondi} M.,  {Ciliegi} P.,  {Schinnerer} E.,  {Smol{\v c}i{\'c}} V.,  {Jahnke}
  K.,  {Carilli} C.,   {Zamorani} G.,  2008, \mn@doi [\apj] {10.1086/589324},
  \href {http://adsabs.harvard.edu/abs/2008ApJ...681.1129B} {681, 1129}

\bibitem[\protect\citeauthoryear{{Bondi} et~al.,}{{Bondi}
  et~al.}{2018}]{bondi18}
{Bondi} M.,  et~al., 2018, \mn@doi [\aap] {10.1051/0004-6361/201834243}, \href
  {https://ui.adsabs.harvard.edu/abs/2018A&A...618L...8B} {618, L8}

\bibitem[\protect\citeauthoryear{{Brammer} et~al.,}{{Brammer}
  et~al.}{2012}]{brammer12}
{Brammer} G.~B.,  et~al., 2012, \mn@doi [\apjs] {10.1088/0067-0049/200/2/13},
  \href {https://ui.adsabs.harvard.edu/abs/2012ApJS..200...13B} {200, 13}

\bibitem[\protect\citeauthoryear{{Brandt} et~al.,}{{Brandt}
  et~al.}{2001}]{brandt2001}
{Brandt} W.~N.,  et~al., 2001, \mn@doi [\aj] {10.1086/324105}, \href
  {https://ui.adsabs.harvard.edu/abs/2001AJ....122.2810B} {122, 2810}

\bibitem[\protect\citeauthoryear{{Bressan}, {Silva}  \& {Granato}}{{Bressan}
  et~al.}{2002}]{bressan02}
{Bressan} A.,  {Silva} L.,   {Granato} G.~L.,  2002, \mn@doi [\aap]
  {10.1051/0004-6361:20020960}, \href
  {https://ui.adsabs.harvard.edu/abs/2002A&A...392..377B} {392, 377}

\bibitem[\protect\citeauthoryear{{Bridle} \& {Schwab}}{{Bridle} \&
  {Schwab}}{1999}]{bridle1999:in}
{Bridle} A.~H.,  {Schwab} F.~R.,  1999, in {Taylor} G.~B.,  {Carilli} C.~L.,
  {Perley} R.~A.,  eds,  Astronomical Society of the Pacific Conference Series
  Vol. 180, Synthesis Imaging in Radio Astronomy II. p.~371

\bibitem[\protect\citeauthoryear{{Briggs}}{{Briggs}}{1995}]{briggs95}
{Briggs} D.~S.,  1995, in American Astronomical Society Meeting Abstracts.
  p.~1444

\bibitem[\protect\citeauthoryear{{Capak} et~al.,}{{Capak}
  et~al.}{2004}]{capak04}
{Capak} P.,  et~al., 2004, \mn@doi [\aj] {10.1086/380611}, \href
  {https://ui.adsabs.harvard.edu/abs/2004AJ....127..180C} {127, 180}

\bibitem[\protect\citeauthoryear{{Casey}, {Narayanan}  \& {Cooray}}{{Casey}
  et~al.}{2014}]{casey14}
{Casey} C.~M.,  {Narayanan} D.,   {Cooray} A.,  2014, \mn@doi [\physrep]
  {10.1016/j.physrep.2014.02.009}, \href
  {https://ui.adsabs.harvard.edu/abs/2014PhR...541...45C} {541, 45}

\bibitem[\protect\citeauthoryear{{Condon}}{{Condon}}{1992}]{condon92}
{Condon} J.~J.,  1992, \mn@doi [\araa] {10.1146/annurev.aa.30.090192.003043},
  \href {https://ui.adsabs.harvard.edu/abs/1992ARA%26A..30..575C} {30, 575}

\bibitem[\protect\citeauthoryear{{Cornwell}}{{Cornwell}}{2008}]{cornwell08}
{Cornwell} T.~J.,  2008, \mn@doi [IEEE Journal of Selected Topics in Signal
  Processing] {10.1109/JSTSP.2008.2006388}, \href
  {https://ui.adsabs.harvard.edu/abs/2008ISTSP...2..793C} {2, 793}

\bibitem[\protect\citeauthoryear{{Dickinson}, {Papovich}, {Ferguson}  \&
  {Budav{\'a}ri}}{{Dickinson} et~al.}{2003}]{dickinson03}
{Dickinson} M.,  {Papovich} C.,  {Ferguson} H.~C.,   {Budav{\'a}ri} T.,  2003,
  \mn@doi [\apj] {10.1086/368111}, \href
  {https://ui.adsabs.harvard.edu/abs/2003ApJ...587...25D} {587, 25}

\bibitem[\protect\citeauthoryear{{Driver} et~al.,}{{Driver}
  et~al.}{2009}]{driver09}
{Driver} S.~P.,  et~al., 2009, \mn@doi [Astronomy and Geophysics]
  {10.1111/j.1468-4004.2009.50512.x}, \href
  {https://ui.adsabs.harvard.edu/abs/2009A%26G....50e..12D} {50, 5.12}

\bibitem[\protect\citeauthoryear{{Galvin} et~al.,}{{Galvin}
  et~al.}{2018}]{galvin18}
{Galvin} T.~J.,  et~al., 2018, \mn@doi [\mnras] {10.1093/mnras/stx2613}, \href
  {https://ui.adsabs.harvard.edu/abs/2018MNRAS.474..779G} {474, 779}

\bibitem[\protect\citeauthoryear{{Giavalisco} et~al.,}{{Giavalisco}
  et~al.}{2004}]{giavalisco14}
{Giavalisco} M.,  et~al., 2004, \mn@doi [\apjl] {10.1086/379232}, \href
  {https://ui.adsabs.harvard.edu/abs/2004ApJ...600L..93G} {600, L93}

\bibitem[\protect\citeauthoryear{{G{\'o}mez-Guijarro}
  et~al.,}{{G{\'o}mez-Guijarro} et~al.}{2019}]{gomez-guijarro19}
{G{\'o}mez-Guijarro} C.,  et~al., 2019, \mn@doi [\apj]
  {10.3847/1538-4357/ab418b}, \href
  {https://ui.adsabs.harvard.edu/abs/2019ApJ...886...88G} {886, 88}

\bibitem[\protect\citeauthoryear{{Graham}, {Driver}, {Petrosian}, {Conselice},
  {Bershady}, {Crawford}  \& {Goto}}{{Graham} et~al.}{2005}]{graham05}
{Graham} A.~W.,  {Driver} S.~P.,  {Petrosian} V.,  {Conselice} C.~J.,
  {Bershady} M.~A.,  {Crawford} S.~M.,   {Goto} T.,  2005, \mn@doi [\aj]
  {10.1086/444475}, \href
  {https://ui.adsabs.harvard.edu/abs/2005AJ....130.1535G} {130, 1535}

\bibitem[\protect\citeauthoryear{{Greisen}}{{Greisen}}{2003}]{Griesen2003:aips}
{Greisen} E.~W.,  2003, in {Heck} A.,  ed.,  Astrophysics and Space Science
  Library Vol. 285, Information Handling in Astronomy - Historical Vistas.
  p.~109, \mn@doi{10.1007/0-306-48080-8_7}

\bibitem[\protect\citeauthoryear{{Grogin} et~al.,}{{Grogin}
  et~al.}{2011}]{grogin11}
{Grogin} N.~A.,  et~al., 2011, \mn@doi [\apjs] {10.1088/0067-0049/197/2/35},
  \href {https://ui.adsabs.harvard.edu/abs/2011ApJS..197...35G} {197, 35}

\bibitem[\protect\citeauthoryear{{Guidetti} et~al.,}{{Guidetti}
  et~al.}{2017}]{Guidetti2017:em}
{Guidetti} D.,  et~al., 2017, \mn@doi [\mnras] {10.1093/mnras/stx1162}, \href
  {http://adsabs.harvard.edu/abs/2017MNRAS.471..210G} {471, 210}

\bibitem[\protect\citeauthoryear{{Hales}, {Murphy}, {Curran}, {Middelberg},
  {Gaensler}  \& {Norris}}{{Hales} et~al.}{2012}]{Hales2012:blo}
{Hales} C.~A.,  {Murphy} T.,  {Curran} J.~R.,  {Middelberg} E.,  {Gaensler}
  B.~M.,   {Norris} R.~P.,  2012, \mn@doi [\mnras]
  {10.1111/j.1365-2966.2012.21373.x}, \href
  {http://adsabs.harvard.edu/abs/2012MNRAS.425..979H} {425, 979}

\bibitem[\protect\citeauthoryear{{Hancock}, {Trott}  \&
  {Hurley-Walker}}{{Hancock} et~al.}{2018}]{Hancock2018:ae}
{Hancock} P.~J.,  {Trott} C.~M.,   {Hurley-Walker} N.,  2018, \mn@doi [\pasa]
  {10.1017/pasa.2018.3}, \href
  {http://adsabs.harvard.edu/abs/2018PASA...35...11H} {35, e011}

\bibitem[\protect\citeauthoryear{{Harrison}, {Costa}, {Tadhunter},
  {Fl{\"u}tsch}, {Kakkad}, {Perna}  \& {Vietri}}{{Harrison}
  et~al.}{2018}]{harrison18}
{Harrison} C.~M.,  {Costa} T.,  {Tadhunter} C.~N.,  {Fl{\"u}tsch} A.,  {Kakkad}
  D.,  {Perna} M.,   {Vietri} G.,  2018, \mn@doi [Nature Astronomy]
  {10.1038/s41550-018-0403-6}, \href
  {https://ui.adsabs.harvard.edu/abs/2018NatAs...2..198H} {2, 198}

\bibitem[\protect\citeauthoryear{{Harrison} et~al.,}{{Harrison}
  et~al.}{2020}]{harrison20}
{Harrison} I.,  et~al., 2020, \mn@doi [\mnras] {10.1093/mnras/staa696}, \href
  {https://ui.adsabs.harvard.edu/abs/2020MNRAS.tmp.1079H} {}

\bibitem[\protect\citeauthoryear{{Helou}, {Soifer}  \&
  {Rowan-Robinson}}{{Helou} et~al.}{1985}]{helou85}
{Helou} G.,  {Soifer} B.~T.,   {Rowan-Robinson} M.,  1985, \mn@doi [\apjl]
  {10.1086/184556}, \href {http://adsabs.harvard.edu/abs/1985ApJ...298L...7H}
  {298, L7}

\bibitem[\protect\citeauthoryear{{Hodge} et~al.,}{{Hodge}
  et~al.}{2013}]{hodge13}
{Hodge} J.~A.,  et~al., 2013, \mn@doi [\apj] {10.1088/0004-637X/768/1/91},
  \href {https://ui.adsabs.harvard.edu/abs/2013ApJ...768...91H} {768, 91}

\bibitem[\protect\citeauthoryear{{Hodge}, {Riechers}, {Decarli}, {Walter},
  {Carilli}, {Daddi}  \& {Dannerbauer}}{{Hodge} et~al.}{2015}]{hodge15}
{Hodge} J.~A.,  {Riechers} D.,  {Decarli} R.,  {Walter} F.,  {Carilli} C.~L.,
  {Daddi} E.,   {Dannerbauer} H.,  2015, \mn@doi [\apjl]
  {10.1088/2041-8205/798/1/L18}, \href
  {https://ui.adsabs.harvard.edu/abs/2015ApJ...798L..18H} {798, L18}

\bibitem[\protect\citeauthoryear{{H{\"o}gbom}}{{H{\"o}gbom}}{1974}]{hogbom74}
{H{\"o}gbom} J.~A.,  1974, \aaps, \href
  {https://ui.adsabs.harvard.edu/abs/1974A%26AS...15..417H} {15, 417}

\bibitem[\protect\citeauthoryear{{Intema}, {van der Tol}, {Cotton}, {Cohen},
  {van Bemmel}  \& {R{\"o}ttgering}}{{Intema} et~al.}{2009}]{intema2009}
{Intema} H.~T.,  {van der Tol} S.,  {Cotton} W.~D.,  {Cohen} A.~S.,  {van
  Bemmel} I.~M.,   {R{\"o}ttgering} H.~J.~A.,  2009, \mn@doi [\aap]
  {10.1051/0004-6361/200811094}, \href
  {https://ui.adsabs.harvard.edu/abs/2009A&A...501.1185I} {501, 1185}

\bibitem[\protect\citeauthoryear{{Ivison} et~al.,}{{Ivison}
  et~al.}{2010}]{ivison10}
{Ivison} R.~J.,  et~al., 2010, \mn@doi [\aap] {10.1051/0004-6361/201014552},
  \href {https://ui.adsabs.harvard.edu/abs/2010A&A...518L..31I} {518, L31}

\bibitem[\protect\citeauthoryear{{Jarvis} et~al.,}{{Jarvis}
  et~al.}{2016}]{jarvis2016}
{Jarvis} M.,  et~al., 2016, in Proceedings of MeerKAT Science: On the Pathway
  to the SKA. 25-27 May. p.~6 (\mn@eprint {arXiv} {1709.01901})

\bibitem[\protect\citeauthoryear{{Jarvis} et~al.,}{{Jarvis}
  et~al.}{2019}]{jarvis19}
{Jarvis} M.~E.,  et~al., 2019, \mn@doi [\mnras] {10.1093/mnras/stz556}, \href
  {https://ui.adsabs.harvard.edu/abs/2019MNRAS.485.2710J} {485, 2710}

\bibitem[\protect\citeauthoryear{{Jim{\'e}nez-Andrade}
  et~al.,}{{Jim{\'e}nez-Andrade} et~al.}{2019}]{jimenez-andrade19}
{Jim{\'e}nez-Andrade} E.~F.,  et~al., 2019, \mn@doi [\aap]
  {10.1051/0004-6361/201935178}, \href
  {https://ui.adsabs.harvard.edu/abs/2019A&A...625A.114J} {625, A114}

\bibitem[\protect\citeauthoryear{{Kajisawa} et~al.,}{{Kajisawa}
  et~al.}{2011}]{kajisawa11}
{Kajisawa} M.,  et~al., 2011, \mn@doi [\pasj] {10.1093/pasj/63.sp2.S379}, \href
  {https://ui.adsabs.harvard.edu/abs/2011PASJ...63S.379K} {63, 379}

\bibitem[\protect\citeauthoryear{{Koekemoer} et~al.,}{{Koekemoer}
  et~al.}{2011}]{koekemoer11}
{Koekemoer} A.~M.,  et~al., 2011, \mn@doi [\apjs] {10.1088/0067-0049/197/2/36},
  \href {https://ui.adsabs.harvard.edu/abs/2011ApJS..197...36K} {197, 36}

\bibitem[\protect\citeauthoryear{{Kroupa}}{{Kroupa}}{2001}]{kroupa01}
{Kroupa} P.,  2001, \mn@doi [\mnras] {10.1046/j.1365-8711.2001.04022.x}, \href
  {https://ui.adsabs.harvard.edu/abs/2001MNRAS.322..231K} {322, 231}

\bibitem[\protect\citeauthoryear{{Lilly}, {Le Fevre}, {Hammer}  \&
  {Crampton}}{{Lilly} et~al.}{1996}]{lilly96}
{Lilly} S.~J.,  {Le Fevre} O.,  {Hammer} F.,   {Crampton} D.,  1996, \mn@doi
  [\apjl] {10.1086/309975}, \href
  {http://adsabs.harvard.edu/abs/1996ApJ...460L...1L} {460, L1}

\bibitem[\protect\citeauthoryear{Lindroos, Knudsen, Stanley, Muxlow, Beswick,
  Conway, Radcliffe  \& Wrigley}{Lindroos et~al.}{2018}]{lindroos18}
Lindroos L.,  Knudsen K.~K.,  Stanley F.,  Muxlow T. W.~B.,  Beswick R.~J.,
  Conway J.,  Radcliffe J.~F.,   Wrigley N.,  2018, \mn@doi [Monthly Notices of
  the Royal Astronomical Society] {10.1093/mnras/sty426}, 476, 3544

\bibitem[\protect\citeauthoryear{{Madau}, {Ferguson}, {Dickinson},
  {Giavalisco}, {Steidel}  \& {Fruchter}}{{Madau} et~al.}{1996}]{madau96}
{Madau} P.,  {Ferguson} H.~C.,  {Dickinson} M.~E.,  {Giavalisco} M.,  {Steidel}
  C.~C.,   {Fruchter} A.,  1996, \mn@doi [\mnras] {10.1093/mnras/283.4.1388},
  \href {https://ui.adsabs.harvard.edu/abs/1996MNRAS.283.1388M} {283, 1388}

\bibitem[\protect\citeauthoryear{{Magnelli} et~al.,}{{Magnelli}
  et~al.}{2015}]{magnelli15}
{Magnelli} B.,  et~al., 2015, \mn@doi [\aap] {10.1051/0004-6361/201424937},
  \href {http://adsabs.harvard.edu/abs/2015A%26A...573A..45M} {573, A45}

\bibitem[\protect\citeauthoryear{{McMullin}, {Waters}, {Schiebel}, {Young}  \&
  {Golap}}{{McMullin} et~al.}{2007}]{mcmullin07}
{McMullin} J.~P.,  {Waters} B.,  {Schiebel} D.,  {Young} W.,   {Golap} K.,
  2007, {CASA Architecture and Applications}.
p.~127

\bibitem[\protect\citeauthoryear{{Mohan} \& {Rafferty}}{{Mohan} \&
  {Rafferty}}{2015}]{mohan15}
{Mohan} N.,  {Rafferty} D.,  2015, {PyBDSF: Python Blob Detection and Source
  Finder} (\mn@eprint {ascl} {1502.007})

\bibitem[\protect\citeauthoryear{{Mooley} et~al.,}{{Mooley}
  et~al.}{2016}]{mooley2016}
{Mooley} K.~P.,  et~al., 2016, \mn@doi [\apj] {10.3847/0004-637X/818/2/105},
  \href {https://ui.adsabs.harvard.edu/abs/2016ApJ...818..105M} {818, 105}

\bibitem[\protect\citeauthoryear{{Morrison}, {Owen}, {Dickinson}, {Ivison}  \&
  {Ibar}}{{Morrison} et~al.}{2010}]{morrison10}
{Morrison} G.~E.,  {Owen} F.~N.,  {Dickinson} M.,  {Ivison} R.~J.,   {Ibar} E.,
   2010, \mn@doi [\apjs] {10.1088/0067-0049/188/1/178}, \href
  {https://ui.adsabs.harvard.edu/abs/2010ApJS..188..178M} {188, 178}

\bibitem[\protect\citeauthoryear{{Murphy} et~al.,}{{Murphy}
  et~al.}{2011}]{murphy11}
{Murphy} E.~J.,  et~al., 2011, \mn@doi [\apj] {10.1088/0004-637X/737/2/67},
  \href {https://ui.adsabs.harvard.edu/abs/2011ApJ...737...67M} {737, 67}

\bibitem[\protect\citeauthoryear{{Murphy}, {Momjian}, {Condon}, {Chary},
  {Dickinson}, {Inami}, {Taylor}  \& {Weiner}}{{Murphy}
  et~al.}{2017}]{murphy17}
{Murphy} E.~J.,  {Momjian} E.,  {Condon} J.~J.,  {Chary} R.-R.,  {Dickinson}
  M.,  {Inami} H.,  {Taylor} A.~R.,   {Weiner} B.~J.,  2017, \mn@doi [\apj]
  {10.3847/1538-4357/aa62fd}, \href
  {http://adsabs.harvard.edu/abs/2017ApJ...839...35M} {839, 35}

\bibitem[\protect\citeauthoryear{{Muxlow} et~al.,}{{Muxlow}
  et~al.}{2005}]{Muxlow2005:hdf}
{Muxlow} T.~W.~B.,  et~al., 2005, \mn@doi [\mnras]
  {10.1111/j.1365-2966.2005.08824.x}, \href
  {http://adsabs.harvard.edu/abs/2005MNRAS.358.1159M} {358, 1159}

\bibitem[\protect\citeauthoryear{{Offringa} \& {Smirnov}}{{Offringa} \&
  {Smirnov}}{2017}]{offringa17}
{Offringa} A.~R.,  {Smirnov} O.,  2017, \mn@doi [\mnras]
  {10.1093/mnras/stx1547}, \href
  {https://ui.adsabs.harvard.edu/abs/2017MNRAS.471..301O} {471, 301}

\bibitem[\protect\citeauthoryear{{Offringa}, {van de Gronde}  \&
  {Roerdink}}{{Offringa} et~al.}{2012}]{offringa2012}
{Offringa} A.~R.,  {van de Gronde} J.~J.,   {Roerdink} J.~B.~T.~M.,  2012,
  \mn@doi [\aap] {10.1051/0004-6361/201118497}, \href
  {https://ui.adsabs.harvard.edu/abs/2012A&A...539A..95O} {539, A95}

\bibitem[\protect\citeauthoryear{Offringa, McKinley, Hurley-Walker
  et~al.}{Offringa et~al.}{2014}]{Offringa2014:ws}
Offringa A.~R.,  McKinley B.,  Hurley-Walker  et~al., 2014, \mn@doi [MNRAS]
  {10.1093/mnras/stu1368}, 444, 606

\bibitem[\protect\citeauthoryear{{Owen}}{{Owen}}{2018}]{owen18}
{Owen} F.~N.,  2018, \mn@doi [\apjs] {10.3847/1538-4365/aab4a1}, \href
  {https://ui.adsabs.harvard.edu/abs/2018ApJS..235...34O} {235, 34}

\bibitem[\protect\citeauthoryear{{Peck} \& {Fenech}}{{Peck} \&
  {Fenech}}{2013}]{peck13}
{Peck} L.~W.,  {Fenech} D.~M.,  2013, \mn@doi [Astronomy and Computing]
  {10.1016/j.ascom.2013.09.001}, \href
  {https://ui.adsabs.harvard.edu/abs/2013A%26C.....2...54P} {2, 54}

\bibitem[\protect\citeauthoryear{{Perley} \& {Butler}}{{Perley} \&
  {Butler}}{2013}]{PerleyButler2013:fl}
{Perley} R.~A.,  {Butler} B.~J.,  2013, \mn@doi [\apjs]
  {10.1088/0067-0049/204/2/19}, \href
  {http://adsabs.harvard.edu/abs/2013ApJS..204...19P} {204, 19}

\bibitem[\protect\citeauthoryear{Petrosian}{Petrosian}{1976}]{petrosian1976surface}
Petrosian V.,  1976, The Astrophysical Journal, 209, L1

\bibitem[\protect\citeauthoryear{{Planck Collaboration} et~al.,}{{Planck
  Collaboration} et~al.}{2018}]{planck18}
{Planck Collaboration} et~al., 2018, arXiv e-prints, \href
  {https://ui.adsabs.harvard.edu/abs/2018arXiv180706209P} {p. arXiv:1807.06209}

\bibitem[\protect\citeauthoryear{{Prandoni} \& {Seymour}}{{Prandoni} \&
  {Seymour}}{2015}]{prandoni15}
{Prandoni} I.,  {Seymour} N.,  2015, in Advancing Astrophysics with the Square
  Kilometre Array (AASKA14). p.~67 (\mn@eprint {arXiv} {1412.6512})

\bibitem[\protect\citeauthoryear{{Radcliffe} et~al.,}{{Radcliffe}
  et~al.}{2018}]{radcliffe18}
{Radcliffe} J.~F.,  et~al., 2018, \mn@doi [\aap] {10.1051/0004-6361/201833399},
  \href {https://ui.adsabs.harvard.edu/abs/2018A&A...619A..48R} {619, A48}

\bibitem[\protect\citeauthoryear{{Radcliffe}, {Beswick}, {Thomson}, {Garrett},
  {Barthel}  \& {Muxlow}}{{Radcliffe} et~al.}{2019}]{radcliffe19}
{Radcliffe} J.~F.,  {Beswick} R.~J.,  {Thomson} A.~P.,  {Garrett} M.~A.,
  {Barthel} P.~D.,   {Muxlow} T. W.~B.,  2019, \mn@doi [\mnras]
  {10.1093/mnras/stz2748}, \href
  {https://ui.adsabs.harvard.edu/abs/2019MNRAS.490.4024R} {490, 4024}

\bibitem[\protect\citeauthoryear{{Richards}}{{Richards}}{2000}]{richards00}
{Richards} E.~A.,  2000, \mn@doi [\apj] {10.1086/308684}, \href
  {https://ui.adsabs.harvard.edu/abs/2000ApJ...533..611R} {533, 611}

\bibitem[\protect\citeauthoryear{{Richards}, {Kellermann}, {Fomalont},
  {Windhorst}  \& {Partridge}}{{Richards} et~al.}{1998}]{Richards1998}
{Richards} E.~A.,  {Kellermann} K.~I.,  {Fomalont} E.~B.,  {Windhorst} R.~A.,
  {Partridge} R.~B.,  1998, \mn@doi [\aj] {10.1086/300489}, \href
  {https://ui.adsabs.harvard.edu/abs/1998AJ....116.1039R} {116, 1039}

\bibitem[\protect\citeauthoryear{{Richards} et~al.,}{{Richards}
  et~al.}{2007}]{richards2007}
{Richards} A.~M.~S.,  et~al., 2007, \mn@doi [\aap]
  {10.1051/0004-6361:20077598}, \href
  {https://ui.adsabs.harvard.edu/abs/2007A&A...472..805R} {472, 805}

\bibitem[\protect\citeauthoryear{{Rujopakarn} et~al.,}{{Rujopakarn}
  et~al.}{2016}]{rujopakarn16}
{Rujopakarn} W.,  et~al., 2016, \mn@doi [\apj] {10.3847/0004-637X/833/1/12},
  \href {https://ui.adsabs.harvard.edu/abs/2016ApJ...833...12R} {833, 12}

\bibitem[\protect\citeauthoryear{{Schaye} et~al.,}{{Schaye}
  et~al.}{2015}]{schaye15}
{Schaye} J.,  et~al., 2015, \mn@doi [\mnras] {10.1093/mnras/stu2058}, \href
  {https://ui.adsabs.harvard.edu/abs/2015MNRAS.446..521S} {446, 521}

\bibitem[\protect\citeauthoryear{{Schwab}}{{Schwab}}{1984}]{schwab84}
{Schwab} F.~R.,  1984, \mn@doi [\aj] {10.1086/113605}, \href
  {https://ui.adsabs.harvard.edu/abs/1984AJ.....89.1076S} {89, 1076}

\bibitem[\protect\citeauthoryear{{Scoville} et~al.,}{{Scoville}
  et~al.}{2007}]{scoville07}
{Scoville} N.,  et~al., 2007, \mn@doi [\apjs] {10.1086/516585}, \href
  {https://ui.adsabs.harvard.edu/abs/2007ApJS..172....1S} {172, 1}

\bibitem[\protect\citeauthoryear{{Serjeant} et~al.,}{{Serjeant}
  et~al.}{2003}]{serjeant03}
{Serjeant} S.,  et~al., 2003, \mn@doi [\mnras]
  {10.1046/j.1365-8711.2003.06862.x}, \href
  {https://ui.adsabs.harvard.edu/abs/2003MNRAS.344..887S} {344, 887}

\bibitem[\protect\citeauthoryear{{Seymour} et~al.,}{{Seymour}
  et~al.}{2008}]{seymour08}
{Seymour} N.,  et~al., 2008, \mn@doi [\mnras]
  {10.1111/j.1365-2966.2008.13166.x}, \href
  {https://ui.adsabs.harvard.edu/abs/2008MNRAS.386.1695S} {386, 1695}

\bibitem[\protect\citeauthoryear{{Skelton} et~al.,}{{Skelton}
  et~al.}{2014}]{Skelton2014;3DHST}
{Skelton} R.~E.,  et~al., 2014, \mn@doi [\apjs] {10.1088/0067-0049/214/2/24},
  \href {http://adsabs.harvard.edu/abs/2014ApJS..214...24S} {214, 24}

\bibitem[\protect\citeauthoryear{{Smail}}{{Smail}}{2002}]{smail02}
{Smail} I.,  2002, \mn@doi [\apss] {10.1023/A:1019536508337}, \href
  {https://ui.adsabs.harvard.edu/abs/2002Ap&SS.281..453S} {281, 453}

\bibitem[\protect\citeauthoryear{{Smol{\v c}i{\'c}} et~al.,}{{Smol{\v c}i{\'c}}
  et~al.}{2009}]{smolcic09}
{Smol{\v c}i{\'c}} V.,  et~al., 2009, \mn@doi [\apj]
  {10.1088/0004-637X/690/1/610}, \href
  {https://ui.adsabs.harvard.edu/abs/2009ApJ...690..610S} {690, 610}

\bibitem[\protect\citeauthoryear{{Smol{\v{c}}i{\'c}}
  et~al.,}{{Smol{\v{c}}i{\'c}} et~al.}{2017}]{smolcic17}
{Smol{\v{c}}i{\'c}} V.,  et~al., 2017, \mn@doi [\aap]
  {10.1051/0004-6361/201628704}, \href
  {https://ui.adsabs.harvard.edu/abs/2017A&A...602A...1S} {602, A1}

\bibitem[\protect\citeauthoryear{{Stach} et~al.,}{{Stach}
  et~al.}{2019}]{stach19}
{Stach} S.~M.,  et~al., 2019, \mn@doi [\mnras] {10.1093/mnras/stz1536}, \href
  {https://ui.adsabs.harvard.edu/abs/2019MNRAS.487.4648S} {487, 4648}

\bibitem[\protect\citeauthoryear{{Swinbank} et~al.,}{{Swinbank}
  et~al.}{2014}]{swinbank14}
{Swinbank} A.~M.,  et~al., 2014, \mn@doi [\mnras] {10.1093/mnras/stt2273},
  \href {https://ui.adsabs.harvard.edu/abs/2014MNRAS.438.1267S} {438, 1267}

\bibitem[\protect\citeauthoryear{{Taylor} \& {Jarvis}}{{Taylor} \&
  {Jarvis}}{2017}]{taylor2017}
{Taylor} A.~R.,  {Jarvis} M.,  2017, in Materials Science and Engineering
  Conference Series. p. 012014, \mn@doi{10.1088/1757-899X/198/1/012014}

\bibitem[\protect\citeauthoryear{{Thomson} et~al.,}{{Thomson}
  et~al.}{2014}]{thomson14}
{Thomson} A.~P.,  et~al., 2014, \mn@doi [\mnras] {10.1093/mnras/stu839}, \href
  {http://adsabs.harvard.edu/abs/2014MNRAS.442..577T} {442, 577}

\bibitem[\protect\citeauthoryear{{Thomson}, {Ivison}, {Owen}, {Danielson},
  {Swinbank}  \& {Smail}}{{Thomson} et~al.}{2015}]{thomson15}
{Thomson} A.~P.,  {Ivison} R.~J.,  {Owen} F.~N.,  {Danielson} A.~L.~R.,
  {Swinbank} A.~M.,   {Smail} I.,  2015, \mn@doi [\mnras]
  {10.1093/mnras/stv118}, \href
  {https://ui.adsabs.harvard.edu/abs/2015MNRAS.448.1874T} {448, 1874}

\bibitem[\protect\citeauthoryear{{Thomson} et~al.,}{{Thomson}
  et~al.}{2017}]{thomson17}
{Thomson} A.~P.,  et~al., 2017, \mn@doi [\apj] {10.3847/1538-4357/aa61a6},
  \href {https://ui.adsabs.harvard.edu/abs/2017ApJ...838..119T} {838, 119}

\bibitem[\protect\citeauthoryear{{Thomson} et~al.,}{{Thomson}
  et~al.}{2019}]{thomson19b}
{Thomson} A.~P.,  et~al., 2019, \mn@doi [\apj] {10.3847/1538-4357/ab32e7},
  \href {https://ui.adsabs.harvard.edu/abs/2019ApJ...883..204T} {883, 204}

\bibitem[\protect\citeauthoryear{{Tisani{\'c}} et~al.,}{{Tisani{\'c}}
  et~al.}{2019}]{tisanic19}
{Tisani{\'c}} K.,  et~al., 2019, \mn@doi [\aap] {10.1051/0004-6361/201834002},
  \href {https://ui.adsabs.harvard.edu/abs/2019A&A...621A.139T} {621, A139}

\bibitem[\protect\citeauthoryear{Tukey}{Tukey}{1962}]{tukey62}
Tukey J.~W.,  1962, \mn@doi [Ann. Math. Statist.] {10.1214/aoms/1177704711},
  33, 1

\bibitem[\protect\citeauthoryear{{Williams} et~al.,}{{Williams}
  et~al.}{1996}]{williams96}
{Williams} R.~E.,  et~al., 1996, \mn@doi [\aj] {10.1086/118105}, \href
  {https://ui.adsabs.harvard.edu/abs/1996AJ....112.1335W} {112, 1335}

\bibitem[\protect\citeauthoryear{{Williams}, {Quadri}, {Franx}, {van Dokkum},
  {Toft}, {Kriek}  \& {Labb{\'e}}}{{Williams} et~al.}{2010}]{williams10}
{Williams} R.~J.,  {Quadri} R.~F.,  {Franx} M.,  {van Dokkum} P.,  {Toft} S.,
  {Kriek} M.,   {Labb{\'e}} I.,  2010, \mn@doi [\apj]
  {10.1088/0004-637X/713/2/738}, \href
  {https://ui.adsabs.harvard.edu/abs/2010ApJ...713..738W} {713, 738}

\bibitem[\protect\citeauthoryear{{Wrigley}}{{Wrigley}}{2016}]{wrigley16}
{Wrigley} N.~H.,  2016, Deep Observations of the GOODS-North Field from the
  $e$MERGE Survey, PhD thesis, Univ. of Manchester

\bibitem[\protect\citeauthoryear{{van der Wel} et~al.,}{{van der Wel}
  et~al.}{2014}]{vanderwel14}
{van der Wel} A.,  et~al., 2014, \mn@doi [\apj] {10.1088/0004-637X/788/1/28},
  \href {https://ui.adsabs.harvard.edu/abs/2014ApJ...788...28V} {788, 28}

\makeatother
\end{thebibliography}

\appendix

\section{Author Affiliations}\label{appendix:affiliations}
$^{1}$Jodrell Bank Centre for Astrophysics, The University of Manchester, SK11\,9DL, UK\\
$^{2}$Department of Physics and Astronomy, University of Southampton, University Road, Southampton, SO17\,1BJ\\
$^{3}$Department of Physics, University of Pretoria, Lynnwood Road, Hatfield, Pretoria 0083, South Africa\\
$^{4}$Centre for Extragalactic Astronomy, Department of Physics, Durham University, South Road, Durham, DH1\,3LE, UK\\
$^{5}$European Southern Observatory, Karl-Schwarzschild-Stra\ss e 2, D-85748 Garching bei M\"unchen, Germany\\
$^{6}$ Institute for Astronomy, University of Edinburgh, Royal Observatory, Blackford Hill, Edinburgh EH9\,3HJ, UK\\
$^{7}$Department of Physics \& Astronomy, University of the Western Cape, Private Bag X17, Belville 7535, South Africa\\
$^{8}$Astrophysics, University of Oxford, Denys Wilkinson Building, Keble Road, Oxford, OX1\,3RH, UK\\
$^{9}$INAF-Instituto di Radioastronomia, Via P.\,Gobetti 101, I-40129, Bologna, Italy\\
$^{10}$University of Central Lancashire, Jeremiah Horrocks Institute, Preston, PR1\,2HE, UK\\
$^{11}$Institute of Cosmology and Gravitation, University of Portsmouth, Portsmouth, PO1\,3FX, UK\\
$^{12}$Department of Physics and Atmospheric Science, Dalhousie University, Halifax, NS B3H\,3J5, Canada\\
$^{13}$Centre for Astrophysics Research, School of Physics, Astronomy \& Mathematics, University of Hertfordshire, Hatfield, AL10\,9AB, UK\\
$^{14}$Shanghai Astronomical Observatory, 80 Nandan Road, Xuihui District, Shanghai, 200030, China\\
$^{15}$University of Chinese Academy of Sciences, 19A Yuquanlu, Beijing, 100049, China\\
$^{16}$Leiden Observatory, Leiden University, P.O.\ Box 9513m 2300 RA Leiden, The Netherlands\\
$^{17}$Instituto de F\'isica y Astronom\'ia, Universidad de Valpara\'iso, Avda. Gran Breta\~na 1111, Valpara\'iso, Chile\\
$^{18}$Institute of Physics, Laboratory of Astrophysics, Ecole Polytechnique F\'{e}d\'{e}rale de Lausanne (EPFL), Observatoire de Sauverny \\ 1290 Versoix, Switzerland\\
$^{19}$Aix Marseille Universit\'{e}, CNRS, LAM (Laboratoire d'Astrophysique de Marseille) UMR 7326, 13388, Marseille, France\\
$^{20}$Department of Space, Earth and Environment, Chalmers University of Technology Onsala Space Observatory, SE-439 92 Onsala, Sweden\\
$^{21}$Kapteyn Astronomical Institute, University of Groningen, P.O.Box 800, 9700AV Groningen, the Netherlands\\
$^{22}$National Radio Astronomy Observatory, 520 Edgemont Road, Charlottesville, VA 22903, USA\\
$^{23}$RAL Space, STFC Rutherford Appleton Laboratory, Didcot, Oxfordshire, OX11\,0QX, UK\\
$^{24}$Instituto de Astrofisica de Andaluc\'ia (IAA, CSIC); Glorieta de la Astronom\'ia s/n, 18008-Granada Spain\\
$^{25}$Astronomy Centre, Department of Physics and Astronomy, University of Sussex, Brighton, BN1\,9QH, UK
$^{26}$Department of Physical Sciences, The Open University, Milton Keynes, MK7\,6AA, UK\\
$^{27}$Gemini Observatory, Northern Operations Center, 670 North A`\={o}h\={o}ku Place, Hilo, HI 96720-2700, USA\\
$^{28}$EACOA fellow: Academia Sinica Institute of Astronomy and Astrophysics, No. 1, Sec. 4, Roosevelt Rd., Taipei 10617, Taiwan\\


\bsp	
\label{lastpage}
\end{document}